\documentclass[PRX,superscriptaddress,11pt]{revtex4-1}
\usepackage{amsmath,amsfonts,amssymb}
\usepackage[T1]{fontenc}
\usepackage{units}
\usepackage{amssymb}
\usepackage{graphicx}
\usepackage{wasysym}
\usepackage{grffile} 
\usepackage{color}
\usepackage{siunitx}
\usepackage{booktabs}

\begin{document}

\title{
Tuning the Ultrafast Response of Fano Resonances in Halide Perovskite Nanoparticles}

\author{Paolo Franceschini}
\email{paolo.franceschini@unicatt.it}
\affiliation{Department of Mathematics and Physics, Universit\`a Cattolica del Sacro Cuore, Brescia I-25121, Italy}
\affiliation{ILAMP (Interdisciplinary Laboratories for Advanced Materials Physics), Universit\`a Cattolica del Sacro Cuore, Brescia I-25121, Italy}
\affiliation{Department of Physics and Astronomy, KU Leuven, Celestijnenlaan 200D, 3001 Leuven, Belgium}

\author{Luca Carletti}
\email{luca.carletti@unibs.it}
\affiliation{Department of Information Engineering, University of Padova, Padova 35131, Italy}
\affiliation{Department of Information Engineering, University of Brescia, Brescia 25123, Italy}

\author{Anatoly P. Pushkarev}
\affiliation{ITMO University, Saint Petersburg 197101, Russia}

\author{Fabrizio Preda}
\affiliation{Dipartimento di Fisica, Politecnico di Milano, Milano 20133, Italy}
\affiliation{NIREOS S.R.L., Via G. Durando 39, 20158 Milano, Italy (www.nireos.com)}

\author{Antonio Perri}
\affiliation{Dipartimento di Fisica, Politecnico di Milano, Milano 20133, Italy}
\affiliation{NIREOS S.R.L., Via G. Durando 39, 20158 Milano, Italy (www.nireos.com)}

\author{Andrea Tognazzi}
\affiliation{Department of Information Engineering, University of Brescia, Brescia 25123, Italy}
\affiliation{National Institute of Optics (INO), Consiglio Nazionale delle Ricerche (CNR), Brescia 25123, Italy}

\author{Andrea Ronchi}
\affiliation{Department of Mathematics and Physics, Universit\`a Cattolica del Sacro Cuore, Brescia I-25121, Italy}
\affiliation{ILAMP (Interdisciplinary Laboratories for Advanced Materials Physics), Universit\`a Cattolica del Sacro Cuore, Brescia I-25121, Italy}
\affiliation{Department of Physics and Astronomy, KU Leuven, Celestijnenlaan 200D, 3001 Leuven, Belgium}

\author{Gabriele Ferrini}
\affiliation{Department of Mathematics and Physics, Universit\`a Cattolica del Sacro Cuore, Brescia I-25121, Italy}
\affiliation{ILAMP (Interdisciplinary Laboratories for Advanced Materials Physics), Universit\`a Cattolica del Sacro Cuore, Brescia I-25121, Italy}

\author{Stefania Pagliara}
\affiliation{Department of Mathematics and Physics, Universit\`a Cattolica del Sacro Cuore, Brescia I-25121, Italy}
\affiliation{ILAMP (Interdisciplinary Laboratories for Advanced Materials Physics), Universit\`a Cattolica del Sacro Cuore, Brescia I-25121, Italy}

\author{Francesco Banfi}
\affiliation{Universit\'e de Lyon, Institut Lumi\`ere Mati\`ere (iLM), Universit\'e Lyon 1 and CNRS 10 rue Ada Byron, 69622 Villeurbanne cedex, France}

\author{Dario Polli}
\affiliation{Dipartimento di Fisica, Politecnico di Milano, Milano 20133, Italy}
\affiliation{NIREOS S.R.L., Via G. Durando 39, 20158 Milano, Italy (www.nireos.com)}

\author{Giulio Cerullo}
\affiliation{Dipartimento di Fisica, Politecnico di Milano, Milano 20133, Italy}

\author{Costantino De Angelis}
\affiliation{Department of Information Engineering, University of Brescia, Brescia 25123, Italy}
\affiliation{National Institute of Optics (INO), Consiglio Nazionale delle Ricerche (CNR), Brescia 25123, Italy}

\author{Sergey V. Makarov}
\affiliation{ITMO University, Saint Petersburg 197101, Russia}

\author{Claudio Giannetti}
\email{claudio.giannetti@unicatt.it}
\affiliation{Department of Mathematics and Physics, Universit\`a Cattolica del Sacro Cuore, Brescia I-25121, Italy}
\affiliation{ILAMP (Interdisciplinary Laboratories for Advanced Materials Physics), Universit\`a Cattolica del Sacro Cuore, Brescia I-25121, Italy}

\begin{abstract}
{The full control of the fundamental photophysics of nanosystems at frequencies as high as few THz is key for tunable and ultrafast nano-photonic devices and metamaterials. Here we combine geometrical and ultrafast control of the optical properties of halide perovskite nanoparticles, which constitute a prominent platform for nanophotonics. The pulsed photoinjection of free carriers across the semiconducting gap leads to a sub-picosecond modification of the far-field electromagnetic properties that is fully controlled by the geometry of the system. When the nanoparticle size is tuned so as to achieve the overlap between the narrowband excitons and the geometry-controlled Mie resonances, the ultrafast modulation of the transmittivity is completely reversed with respect to what is usually observed in nanoparticles with different sizes, in bulk systems and in thin films. The interplay between chemical, geometrical and ultrafast tuning offers an additional control parameter with impact on nano-antennas and ultrafast optical switches.  }
\end{abstract}

\maketitle
\section*{Introduction}
The study of light-matter interaction in halide perovskites (HPs) has revolutionized the photonics community in the last decade \cite{review10years,Fu2019}. The wide interest on this class of materials, which comprises compounds of the form ABX$_3$, where A stands for organic (\textit{e.g.}, CH$_3$NH$_3$=MA) or inorganic (\textit{e.g.}, Cs) cations, B is usually lead (Pb), and X is chosen between the halogens I, Br, or Cl, mainly stems from their outstanding optoelectronic properties, such as strong excitonic resonances at room temperature \cite{Green2014}, tunable band-gaps and emission wavelengths within the entire visible spectrum \cite{Jang2015,Protescu2015,Byun2017}, and long carrier diffusion lengths \cite{Wehrenfennig2014,Dong-science-2015}. These features, combined with low-cost fabrication methods, boosted the development of photovoltaic solar cells with power conversion efficiency exceeding 20\% \cite{yang-science-2015}.

\begin{figure*}[t!]
\includegraphics[keepaspectratio,clip,width=\textwidth]{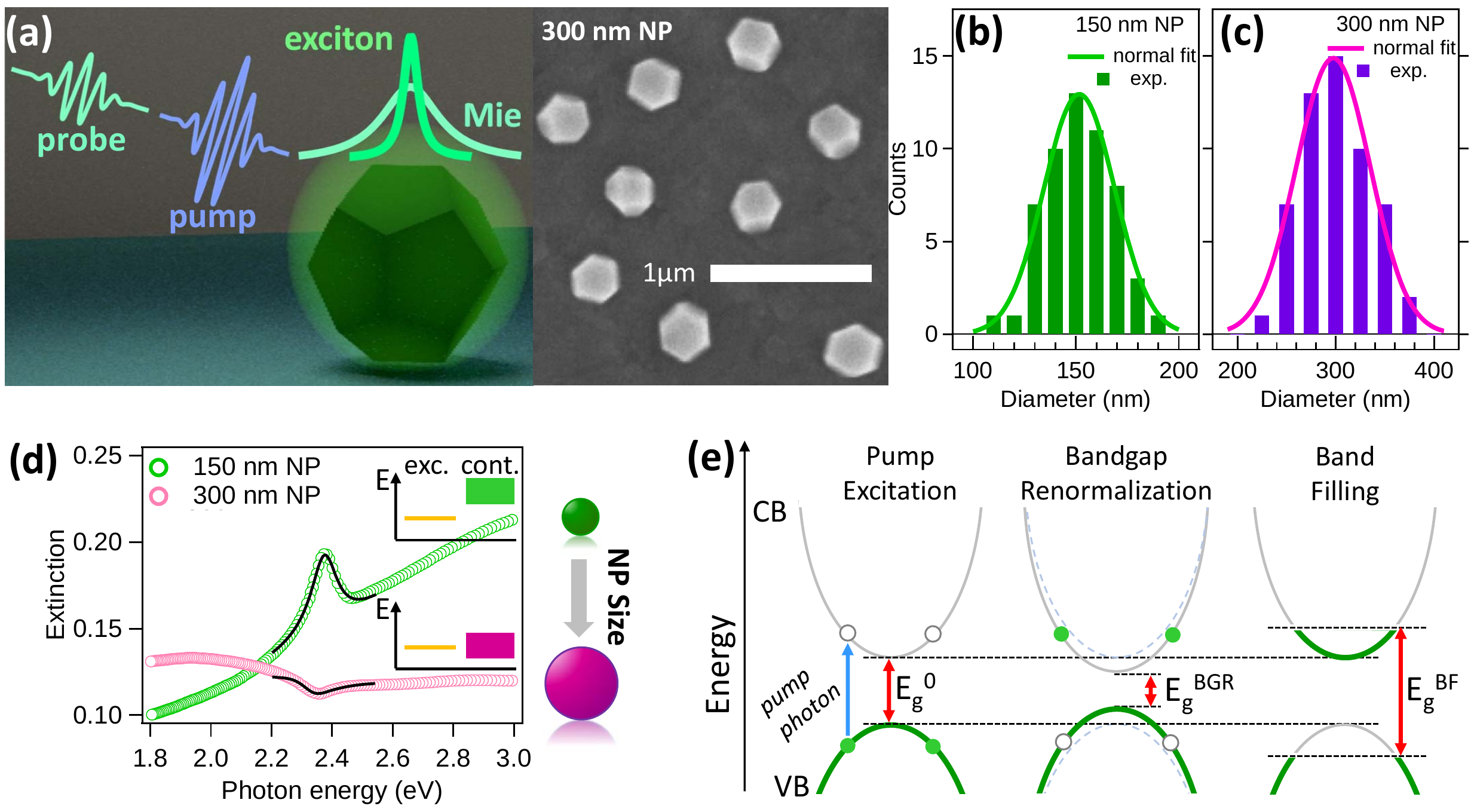}
\caption{\textbf{Structural and Optical Properties of Hybrid Perovskites Nanoparticles.} (a) Concept of the pump-probe experiment on HP nanoparticles. The right panel displays a scanning electron microscopy image of the 300 nm NPs sample. (b, c) Size distribution of CsPbBr$_3$ nanoparticles with average diameter $\bar{\phi}$=150 nm (b, green bars) and 300 nm (c, purple bars). The size distribution has been calculated by analysis of the scanning electron microscopy images. The solid line represents the normal fit to the experimental distributions. For both samples the standard deviation of the normal distribution is 10$\%$ of the average size. (d) Experimental measured extinction spectra of the CsPbBr$_3$ nanoparticles with average diameter of 150 nm (green markers) and 300 nm (pink markers). In the spectral region investigated in the time-resolved experiment, the extinction data have been fitted (solid black lines) with Eq. \ref{eqn:Fano_Resonance}. For the $\SI{150}{nm}$ sample, we estimated $q$=$(-18.1 \pm 0.2)$ and $\Gamma=(0.107 \pm 0.003) \, \SI{}{\eV}$, while $q$=$(-0.41 \pm 0.01)$ and $\Gamma$=$(0.123 \pm 0.005) \, \SI{}{\eV}$ for the $\SI{300}{nm}$ sample. The right panel sketches the energy overlap between the discrete excitonic level (yellow line) and the continuum of states (square), which gives rise to the Fano low-$q$ resonance for 300 nm particles. (e) Schematics of the electron dynamics triggered by the pulsed photo-excitation (blue arrow): the initial creation of the non-thermal electron-hole distributions within the valence (VB) and conduction bands (CB) leads to the renormalization of the equilibrium gap, $E^0_g$, which  reduces to $E_g$=$E^0_g$-$\delta E^{BGR}_g$. At later times, the interactions drive the formation of a long-lived quasi-thermal distribution at the gap edges and a consequent increase of the effective optical band gap, $E_g=E^0_g+\delta E_g^{BF}$. 
}
\label{Fig_structureSize}
\end{figure*}

Halide perovskites constitute also a promising platform for ultrafast optical switching applications. Sub-picosecond visible light pulses can be used to photo-inject free carriers across the HP semiconducting gap, thus triggering the kind of multi-step dynamics that is at the basis of any ultrafast photonic device operating at frequencies as high as several THz \cite{Manser2014,Sheng2015,Milot2015,Yang2015,Price2015,Herz2016,Tamming2019,Palmieri2020}.  Tuning the density and energy-distribution of the photo-carriers thus provides an additional control parameter for achieving the complete tunability of HPs optical properties on the picosecond timescale.

Recent advances in the development of HP-based metamaterials \cite{Gholipour2017,Makarov2017,Gao2018,Berestennikov2019} and nanoantennas \cite{Tiguntseva2018,Makarov2019} brought into play an additional degree of freedom to control HP optical properties on ultrafast timescales \cite{Manjappa2017,Chanana2018,Huang1018}. In this framework, a geometry-based approach was proposed after the observation of tunable Fano resonances in nanoparticles (NPs), arising from the coupling of the discrete excitonic states to the continuum of the geometry-driven Mie modes of the nanostructures  \cite{Makarov2018}. Tuning the interference by a suitable choice of the nanoparticle size allows to dramatically modify the excitonic lineshape, up to the point of reversing the scattering resonant peak into a dip \cite{Makarov2018}.

In this work, we combine geometrical and ultrafast manipulation of the optical conductivity of Mie-resonant CsPbBr$_3$ nanoparticles (see Fig. \ref{Fig_structureSize}a). Our time- and energy-resolved pump-probe data show that, when the nanoparticle size is tuned to achieve a dip in the excitonic Fano lineshape, the photo-induced \textit{band gap renormalization} and the subsequent \textit{band filling} drive a modification of the optical constants, which is opposite as  compared to that observed in bulk HP and in NPs exhibiting a peak in the excitonic Fano lineshape. The analysis of the photophysics of Fano resonances in HP NPs is complemented by finite element simulations, which offer insights into the interplay between the non-equilibrium charge distribution and the coupling of the optical transitions to the geometry-controlled cavity modes and allow us to  extract the intrinsic band-filling dynamics. The combination of optical and geometrical control offers an additional platform for ultrafast optoelectronic nanodevices and metamaterials. CsPbBr$_3$ nanoparticles, exhibiting tunable Fano resonances, represent a prospective building-block for ultrafast all-optical switching in non-linear nanophotonic designs\cite{Shcherbakov2015,Maragkou2015}. The observation of an opposite photo-induced response in  nanoparticles exhibiting different Fano lineshapes represents an interesting method to control the non-linear response of dielectric media in nanodevices and metamaterials with sub-wavelength spatial resolution\cite{Soukoulis2011,Peruch2017}.

\begin{figure*}[h!]
\includegraphics[keepaspectratio,clip,width=0.7\textwidth]{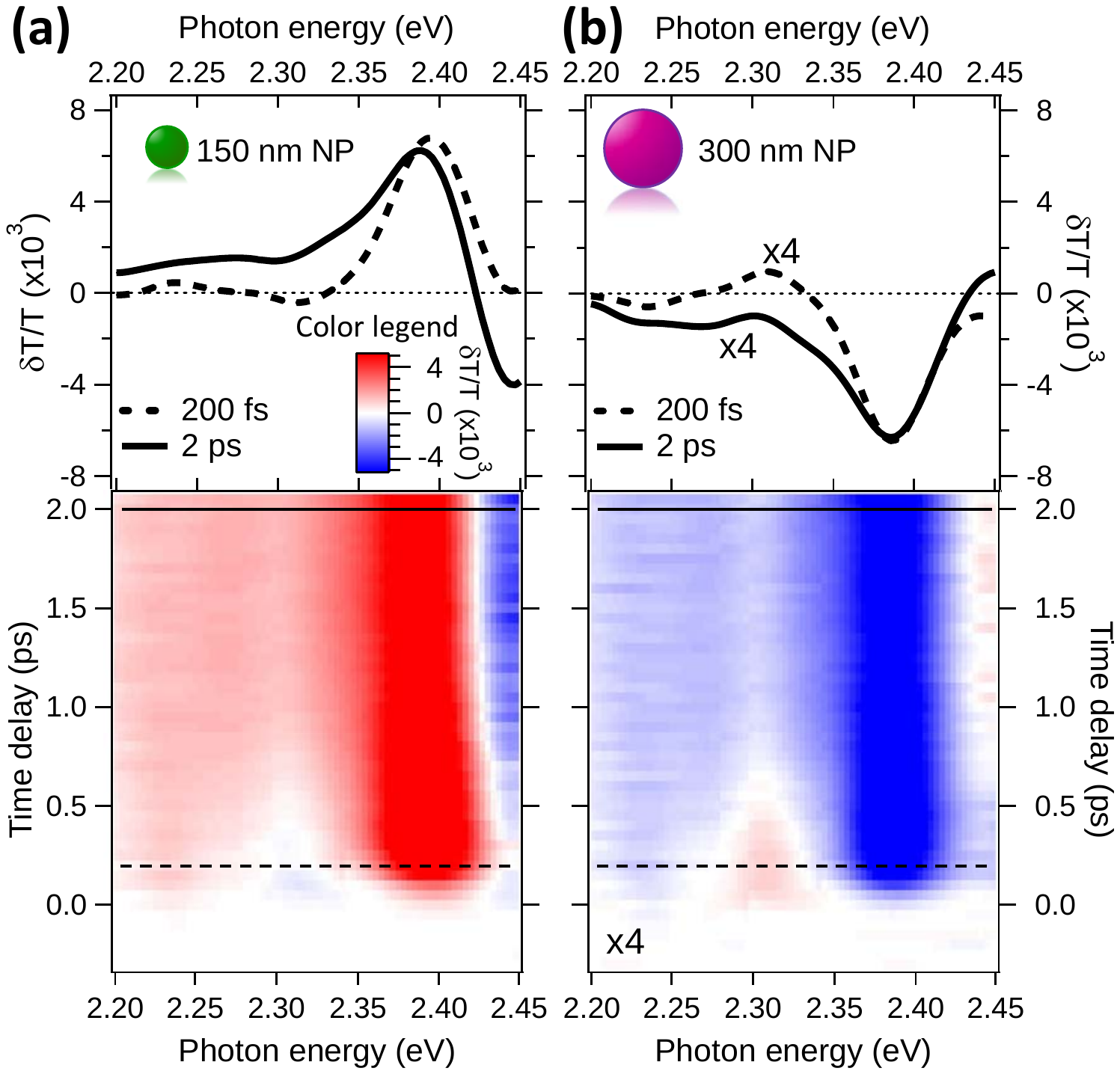}
\caption{\textbf{Ultrafast dynamics of the nanoparticle optical properties.} Energy- and time-resolved differential transmission maps for 150 nm (a) and 300 nm (b) CsPbBr$_3$ nanoparticles pumped at $\SI{2.9}{eV}$, with an incident fluence of $\SI{54}{}\,\mu\mbox{J}$/cm$^2$. The top panels display $\delta  T/T\!\left( \hbar \omega \right)$ spectra at fixed probe delays (200 fs, dashed line; 2 ps solid line) for 150 nm and 300 nm nanoparticles, respectively. For graphical reasons, the data reported in panel (b) are magnified by a factor 4.}
\label{fig:measurements}
\end{figure*}

\section*{Results}
\sloppy
Isolated CsPbBr$_3$ nanoparticles with different average sizes (see distributions in Fig. \ref{Fig_structureSize}b, c) were deposited on a quartz transparent substrate with an average coverage density $\sigma_{cov}$=3 NP/$\mu$m$^2$ \cite{Makarov2018}.
The samples were optically characterized by measuring the extinction, $X$=$-\log_{10} T$, where $T$ is the sample transmission. In Fig. \ref{Fig_structureSize}d we report the extinction, which includes both the scattering and absorption contributions, of samples with 150 and 300 nm average NP diameter ($\bar{\phi}$).  
In both cases, the extinction can be reproduced by the following expression:
\begin{equation} \label{eqn:Fano_Resonance}
X \! \left( \hbar\omega \right) =a \cdot \frac{{\left[ q \cdot \dfrac{\Gamma}{2}+\hbar\omega-(E_g-E_b) \right]}^2}{{ \dfrac{\Gamma^2}{4}}+{\left[ \hbar\omega -(E_g-E_b)  \right]}^2}+X_{\mathrm{bck}}(\hbar\omega,E_g),
\end{equation} 
where the first term on the right-hand side represents a Fano asymmetric excitonic linewidth with amplitude $a$, centered at $E_g-E_b$, $E_b$ being the exciton binding energy, and with a lineshape controlled by the broadening parameter $\Gamma$ and the profile index $q$. $X_{\mathrm{bck}}(\hbar\omega,E_g)$ accounts for the absorption across the semiconducting edge at $E_g$ (see Sec. S2 for more details).


For 150 nm NPs, the small overlap between the excitonic line at $E_g$-$E_b \simeq$2.4 eV and the Mie resonances results in a moderate asymmetry of the peak in the extinction spectrum (green markers in Fig. \ref{Fig_structureSize}d), corresponding to $q\simeq$-18 (\emph{high-$q$ resonance}). When the NP size is increased (300 nm), the overlap between the Mie modes and the excitonic resonance increases. As a result, the scattering contribution strongly increases\cite{Makarov2018} thus turning the extinction lineshape into a dip (pink markers in Fig. \ref{Fig_structureSize}d), which corresponds to $q\simeq$-0.4 (\emph{low-$q$ resonance}). The separate scattering and absorption contributions to the total extinction of single NPs are discussed in the Supplementary Information.

In this work we use $\sim$40 fs (FWHM) light pulses at 2.9 eV photon energy to trigger the out-of-equilibrium dynamics in HP NPs. In general, it is known that in bulk semiconductors the sudden photo-generation of high-density non-thermal electrons(holes) within the conduction(valence) band (see Fig. \ref{Fig_structureSize}e) modifies the electron screening thus shrinking the band gap \cite{Bennett1990}. This phenomenon, known as \textit{band gap renormalization} (BGR), manifests itself into a photo-induced absorption increase below the semiconducting band edge \cite{Price2015}, corresponding to a photo-induced dynamical decrease of the effective band gap, $E_g$=$E^0_g$-$\delta E^{BGR}_g$. Within $\approx$1 ps, the electron-electron and electron-phonon interactions drive the relaxation of the non-thermal population and the creation of a long-lived quasi-thermal distribution, described by an effective temperature and chemical potential, which overfills the electron(hole) states at the band edges (see Fig. \ref{Fig_structureSize}e). The consequent bleaching of the band-edge transitions, known as \textit{band filling} (BF) or \textit{dynamic Burstein-Moss effect} \cite{Burstein1954,Moss1954}, leads to the photo-induced dynamical increase of the effective band gap, $E_g$=$E^0_g$+$\delta E^{BF}_g$, until the equilibrium distribution is eventually recovered on the nanosecond timescale \textit{via} interband recombination \cite{Manser2014}. 
To address the role of the geometry in controlling the modulation of the optical properties consequent to the BGR and BF processes, we performed broadband pump-probe experiments on NPs all the way from Fano high-$q$ to low-$q$ resonance conditions. The pump-induced relative transmittivity variation, $\delta T/T(\hbar \omega, \Delta t)$, is measured in the 2.2-2.5 eV energy range by means of a delayed ($\Delta t$) supercontinuum white light probe detected through a collinear interferometer (GEMINI by NIREOS) \cite{Preda2016,Preda2017}, as described in the Methods. The energy- and time-dependent $\delta T/T(\hbar\omega, \Delta t)$ is reported for 150 and 300 nm NPs (see Figs. \ref{fig:measurements}a, b). More experimental data for different NP sizes are shown in Fig. S1.

In the Fano high-$q$ resonance condition (150 nm, Fig. \ref{fig:measurements}a), the $\delta T/T(\hbar\omega, \Delta t)$ signal is characterized by a short-lived negative component at $\hbar\omega\simeq$2.3 eV and a long-lived signal, which turns from positive to negative at the exciton peak at $E_g$-$E_b\simeq$2.4 eV. The presence of two different spectral components can be appreciated by plotting $\delta T/T$ spectra at different delays ($\Delta t$=0.2 and 2 ps, top panel of Fig. \ref{fig:measurements}a). The negative transmittivity variation at short delays reflects a photo-induced increase of below-gap absorption, which is the signature of the BGR effect, as already observed in thin films \cite{Price2015}. On the picosecond timescale this effect decays as a result of hot carrier cooling through electron-phonon interactions \cite{Price2015}. As a consequence, after $\sim$2 ps the positive(negative) transmittivity variation at $\hbar\omega<$($>$)2.4 eV can be reproduced, over the entire frequency range, by a rigid blue-shift, by $\delta E_g$=(5.1$\pm$0.4) meV, of the excitonic line in Eq. \ref{eqn:Fano_Resonance}. Such shift is the typical manifestation of the conventional \emph{band filling} effect, which has been already reported in HP thin-films\cite{Manser2014}. We remark that the transmittivity variation measured in our pump-probe experiment refers to the \textit{effective} response of the sample, which can be modeled as an inhomogeneous film with CsPbBr$_3$ NPs, with volume filling fraction $f_{vol}$. Therefore, in order to quantitatively compare our results to previous data on thin films, we extracted the pump-induced shift of the band-gap of the individual nanoparticles ($\delta E^{NP}_g$) by calculating the scaling factor ($\tilde{C}$), which relates $\delta E^{NP}_g$ to the measured effective $\delta E_g$. As described in Section S2, we adopted the Modified Maxwell-Garnett Mie (MMGM) theory \cite{Battie2014} for the effective medium, which accounts for size-dispersed spherical particles embedded in a host matrix, to calculate the scaling factor as the ratio between the intrinsic absorption variation of an individual NP and the effective absorption variation of the film. From our analysis, we estimated $\tilde{C}$=60$\pm$40. Considering that, for small shifts, the transmittivity variation in the proximity of a peak is directly proportional to the shift amplitude, we can thus assume that $\delta E^{NP}_g\simeq\tilde{C}\delta E_g$=(300$\pm$200) meV, which is in quantitative agreement with data reported in Ref. \citenum{Manser2014}.

The non-equilibrium optical response of CsPbBr$_3$ nanoparticles progressively changes when the NP size is increased and the overlap between the exciton resonance at $\sim$2.4 eV and the geometry-controlled continuum of the Mie modes is enhanced. In the Fano low-$q$ resonance condition, obtained for 300 nm NP size, the $\delta T/T(\hbar\omega, \Delta t)$ signals associated to both the \textit{band gap renormalization} and \textit{band filling} effects are reversed, as shown in Fig. \ref{fig:measurements}b. In this condition, the sub-ps below-gap transmittivity change is positive, thus suggesting a photo-induced \textit{decrease} of below-gap absorption in opposition to the conventional \textit{band gap renormalization} effect measured in films and nanoparticles far from the Fano low-$q$ resonance condition. 

\begin{figure*}[t]
\includegraphics[keepaspectratio,clip,width=0.9\textwidth]{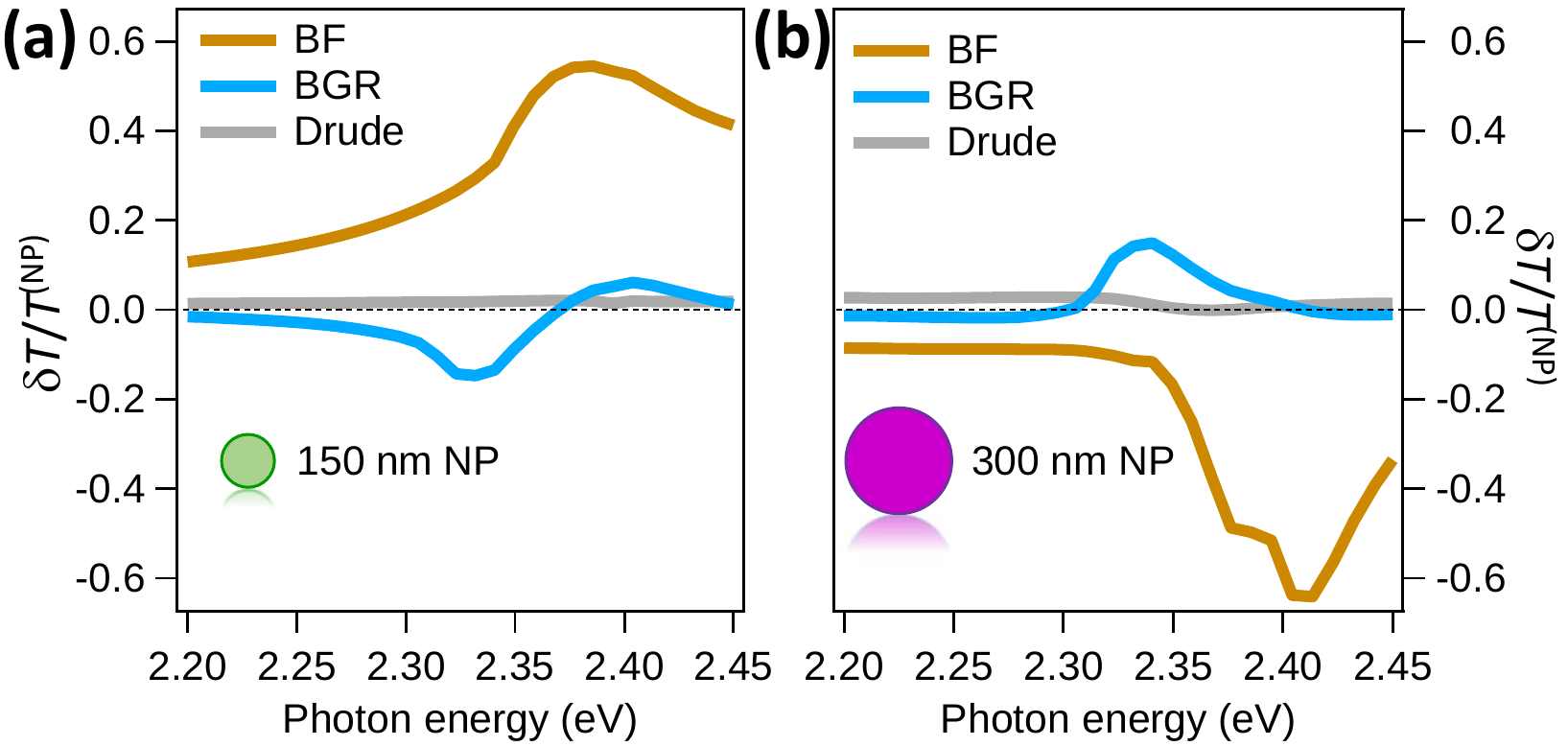}
\caption{\textbf{Numerical simulations.} Differential transmittivity spectrum, $\delta T/T^{\mathrm{(NP)}} \!\left( \hbar \omega \right)$, calculated for individual (a) $\SI{150}{nm}$ and (b) $\SI{300}{nm}$ nanoparticles. The separate contributions of the \emph{bandgap renormalization} (BGR), \emph{band filling} (BF) and \emph{Drude} effects to the total signal are calculated by assuming the photo-injected free carrier density $n_{fc} \simeq \SI{1.2e20}{cm^{-3}}$, for both nanoparticle sizes.
}
\label{fig:simulations}
\end{figure*} 

Our results demonstrate that the non-equilibrium optical properties of CsPbBr$_3$ NPs following ultrafast excitation are crucially controlled by the NP size. The photo-generated electron/hole population in the conduction and valence bands gives rise to a photo-induced variation of the far-field optical properties, which qualitatively and quantitatively depends on the interference between the excitonic resonance and the geometrically-controlled Mie modes. 
To gain further insights into the interplay between non-equilibrium photo-injected free carriers and the geometry-controlled Fano interference, we performed full-vectorial numerical electromagnetic simulations implemented with the finite element method in COMSOL. In these numerical calculations, we considered isolated HP nanospheres deposited on a quartz substrate with refractive index 1.45. The numerical simulations of the full electromagnetic problem go beyond the Mie theory for isolated NPs, which is presented in Section S6, and allow us to account for the NP-substrate interaction that affects  spectral position, amplitude, and phase of the resonances in the NP.~\cite{markovich2014magnetic}

The frequency dependent transmittivity variation of the single NP is calculated as $\delta T/T^{\mathrm{(NP)}}(\hbar\omega)$=$10^{-\left( X^{\mathrm{NP}}_{out} \left( \hbar \omega \right)-X^{\mathrm{NP}}_{eq} \left( \hbar \omega \right) \right)}-1$ (for details, see Methods).  The equilibrium extinction, $X^{\mathrm{NP}}_{eq} \left( \hbar \omega \right)$, is obtained by solving the NP scattering problem and assuming the complex refractive index of bulk CsPbBr$_3$ (extracted from Ref. \citenum{Makarov2018}). The out-of-equilibrium extinction,  $X^{\mathrm{NP}}_{out} \left( \hbar \omega \right)$, is obtained by properly modifying the equilibrium optical properties to account for the effect of  photo-generated non-equilibrium carriers. More specifically, we assume that the photo-excitation process initially injects an excess density of electrons ($n_e$) and holes ($n_h$) approximately equal to the density of absorbed photons, \textit{i.e.} $n_e\simeq n_h\simeq n_{ph}$. On the sub-picosecond timescale, the photo-injected free carrier density, $n_{fc}$=$n_e$+$n_h\simeq$2$n_{ph}\simeq \SI{1.2e20}{cm^{-3}}$ (see Methods for calculation of $n_{ph}$), overcomes the screening critical concentration, $n_{cr}\approx$3.4$\times$10$^{18}$ cm$^{-3}$ (see Sec. S5 of the Supplementary Information)  \cite{Bennett1990}, thus leading to a photo-induced \emph{bandgap renormalization} and a related change of the absorption coefficient expressed as \cite{Bennett1990}:
\begin{eqnarray} \label{eqn:BGR_absorption1}
\delta  \alpha_{BGR} \! \left(\hbar\omega; n_{fc}/n_{cr} \right)=\alpha_{eq}\! \left(\hbar\omega; E^0_g-\delta E_{{BGR}}\! \left(n_{fc}/n_{cr} \right) \right)-\alpha_{eq}\! \left(\hbar\omega; E^0_g \right).
\end{eqnarray} 

Within $\approx$1 ps, the carrier-carrier and carrier-phonon interactions lead to the relaxation of the photo-injected free carriers and to the onset of a quasi-thermal effective carriers distribution which completely fills the states at the bottom(top) of the conduction(valence) band. The bleaching of the gap-edge transitions (\emph{band filling}) manifests itself as  \cite{Bennett1990}:
\begin{eqnarray}\label{eqn:absorption_variation1}
\delta  \alpha_{BF} \! \left( \hbar\omega; n_{fc},T^{*} \right)=
\alpha_{eq}\! \left( \hbar\omega; E^0_g \right) \, \left[ f_{v} \! \left(\hbar\omega;  E^*_{F_{v}}, T^{*} \right)-f_{c} \! \left( \hbar\omega; E^*_{F_{c}}, T^{*} \right) -1 \right],
\end{eqnarray}
where the effective Fermi-Dirac populations in the valence and conduction bands, $f_v$ and $f_c$ respectively, are controlled by the effective Fermi levels $E^*_{F_{v}}$ and $E^*_{F_{c}}$ and by the effective temperature $T^*$. The photo-induced absorption change related to transition of free carriers within the conduction(valence) band is accounted for by a \emph{Drude} model, as described in Sec. S5 of the Supplementary Information.
We calculated $\delta T/T^{\mathrm{(NP)}}(\hbar\omega)$ for different NP sizes, namely 150 and 300 nm (Fig. \ref{fig:simulations}a and b, respectively), corresponding to the Fano high-$q$ resonance and low-$q$ resonance conditions. While the \textit{Drude} contribution is always negligible (see Fig. \ref{fig:simulations}), the BGR and BF dominate the transmittivity change, as can be evidenced by calculating separately the two contributions to the $\delta T/T^{\mathrm{(NP)}}(\hbar\omega)$ signal. For both NP sizes, the BGR effect has been estimated by assuming $n_{fc}/n_{cr}\simeq 2n_{ph}/n_{cr}$=35 in Eq. \ref{eqn:BGR_absorption1}. The BF effect has been reproduced by assuming that the effective thermal energy of the free carriers is given by the excess energy injected by the pump photons and equally shared between electrons and holes, \textit{i.e.} 3/2$kT^*$=($\hbar\omega_p$-$E^0_g$)/2. Considering $\hbar\omega_p$=2.9 eV and $E^0_g\simeq$2.4 eV, we obtain $T^*\simeq$3000 K. $E^*_{F_{v}}$ and $E^*_{F_{c}}$ are then self-consistently calculated (see Eqs. S21 and S22) assuming $n_{fc}$=2$n_{ph}$. The results reported in Fig. \ref{fig:simulations} show that, for 150 nm NP size, the BGR results in an additional below edge absorption (negative $\delta T/T$), which is opposite to the signal related to the BF process. In contrast, when the NP size matches the Fano low-$q$ resonance condition (300 nm, Fig. \ref{fig:simulations}b), both the BGR and BF contributions to the differential transmittivity variation are of opposite sign as compared to the high-$q$ resonant case. Moreover, the results obtained from the simulations are consistent with the experimental data obtained for a wider set of CsPbBr$_3$ NP samples. Indeed, in the case of nanoparticles with average sizes of $\sim \SI{100}{nm}$ and $\sim \SI{220}{nm}$, whose extinction spectrum is characterized by a peak (high-$q$) near the exciton line (see Fig. S1a), their out-of-equilibrium optical properties (reported in Fig. S1b, d) exhibit spectral features similar to those reported for $\SI{150}{nm}$ NP sample. These numerical results support our experimental findings and demonstrates that the photo-induced modulation of the NP optical properties is controlled by the geometry of the system. 

\begin{figure*}[t]
\includegraphics[keepaspectratio,clip,width=0.49\textwidth]{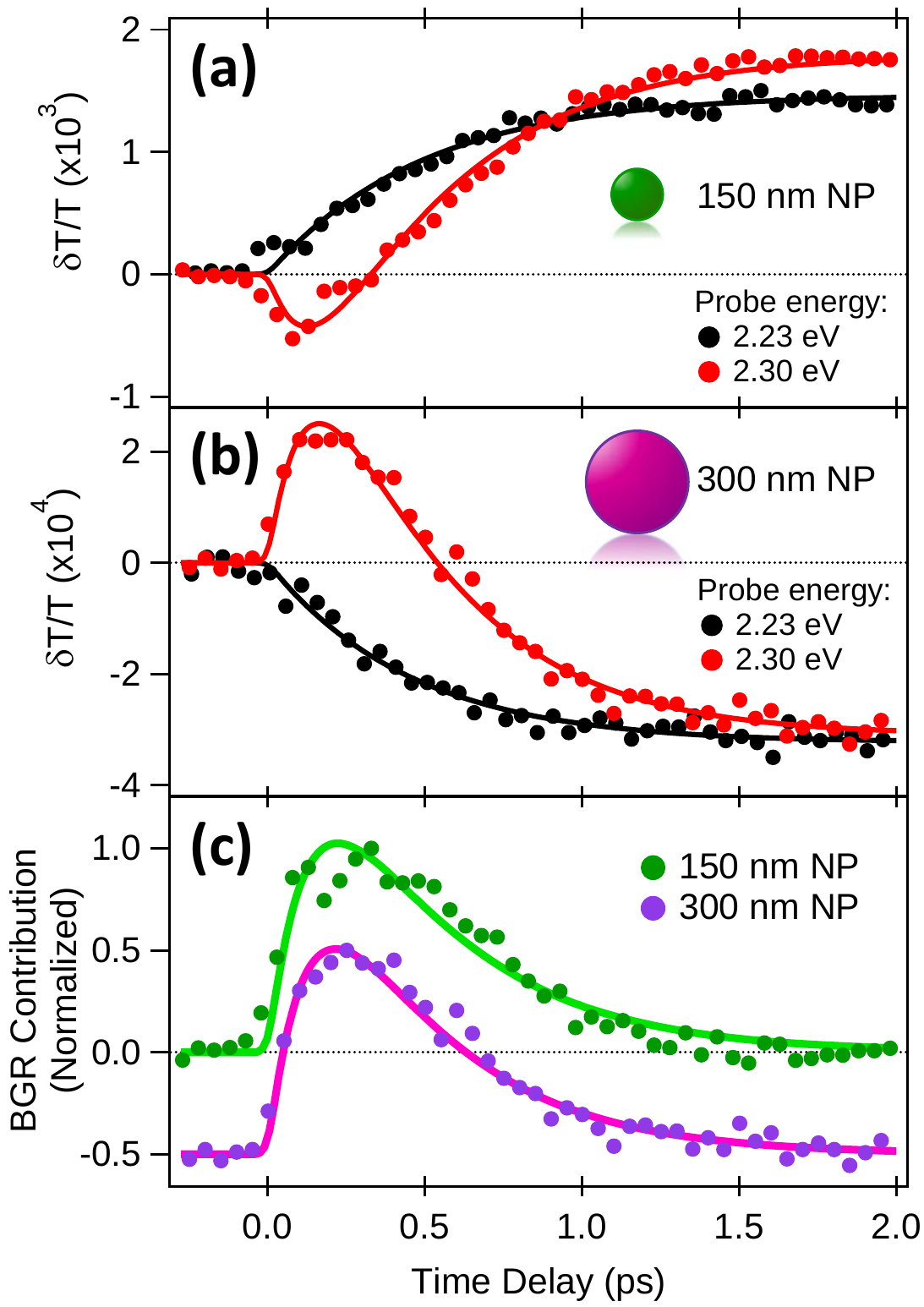}
\caption{\textbf{Disentangling band filling and bandgap renormalization.} Differential time traces at specific photon energies are reported for NP sizes matching the Fano high-$q$ resonance (150 nm, panel (a)) and low-$q$ resonance conditions (300 nm, panel (b)). The black solid lines represent the single exponential fitting to the time traces taken at $\hbar\omega$=2.23 eV (black dots). The red solid lines represent the multi exponential fitting to the time traces taken at $\hbar\omega$=2.3 eV (red dots). Panel (c) displays the BGR signal obtained by subtracting the BF dynamics from the time traces taken at $\hbar\omega$=2.3 eV and shown in panels (a) and (b). The solid lines represent the exponential fitting to the BGR dynamics.
}
\label{fig:fit_dynamics}
\end{figure*} 

The numerical results reported in Fig. \ref{fig:simulations} offer important insights into the photophysics of HP NPs and provide a guide for disentangling the BGR and BF dynamics, which experimentally overlap on the picosecond timescale. The results reported in Figs. \ref{fig:simulations}a and \ref{fig:simulations}b indicate that, for both NP sizes, the $\delta T/T^{\mathrm{(NP)}}$ at $\hbar\omega\simeq$2.2 eV is dominated by the BF effect, whereas the BGR signal is negligible. As a consequence, time traces at this photon energy, although smaller in amplitude with respect to data at higher photon energy, directly reproduce the dynamics of the BF alone. In Figs. \ref{fig:fit_dynamics}a and \ref{fig:fit_dynamics}b we report $\delta T/T(\hbar\omega$=2.23 eV, $\Delta t$) time traces (black dots), which can be fitted by a single exponential function (black lines) with similar time constant $\tau_{\mathrm{BF}}$=480$\pm$20 fs. The BGR dynamics can then be retrieved by analyzing the time-traces taken at different photon energies for which the BGR contribution is not negligible. In Fig. \ref{fig:fit_dynamics}a and \ref{fig:fit_dynamics}b we report $\delta T/T(\hbar\omega$=2.3 eV, $\Delta t$) time traces (red dots), which exhibit a multi-exponential behaviour. In Fig. \ref{fig:fit_dynamics}c we report the components of the multi-exponential fit which directly represent the time evolution of the actual BGR process; the curves are obtained by subtracting the properly scaled BF contribution from the full $\delta T/T$ dynamics at $\hbar\omega=\SI{2.3}{eV}$ (see Sec. S4 for more details). For both NP sizes, the build-up time of the BGR signal is $\approx$200 fs, whereas the relaxation of the non-thermal carriers responsible for the BGR takes place within 400$\pm$10 fs, a timescale compatible with the carrier cooling mediated by the coupling to optical phonons \cite{Richter2017}. We underline that, in the time-resolved traces, the dynamics of the $\delta T/T$ signal is proportional to the non-equilibrium population within the valence and conduction bands. This population, which is bottlenecked by the presence of the gap, decays on a time scale much longer than the inverse of the exciton linewidth ($\sim {\hbar}/{ \Gamma}=\SI{6}{fs}$) that accounts for the whole of quasi-elastic scattering processes which reduce the exciton lifetime.
The procedure here introduced and based on time- and frequency-resolved optical measurements constitutes a way to disentangle the BGR and BF processes. Our results show that, microscopically, the two processes are almost independent of the NP size, as expected since the exciton Bohr's radius ($\sim$7 nm) \cite{Protescu2015} is much smaller than the NP size. The observed inversion of the $\delta T/T(\hbar\omega$, $\Delta t$) signals in the Fano low-$q$ resonance condition is thus a genuine effect of the geometry of the system, which controls the sub-picosecond modulation of the optical properties.

\begin{figure*}[t]
\includegraphics[keepaspectratio,clip,width=0.8\textwidth]{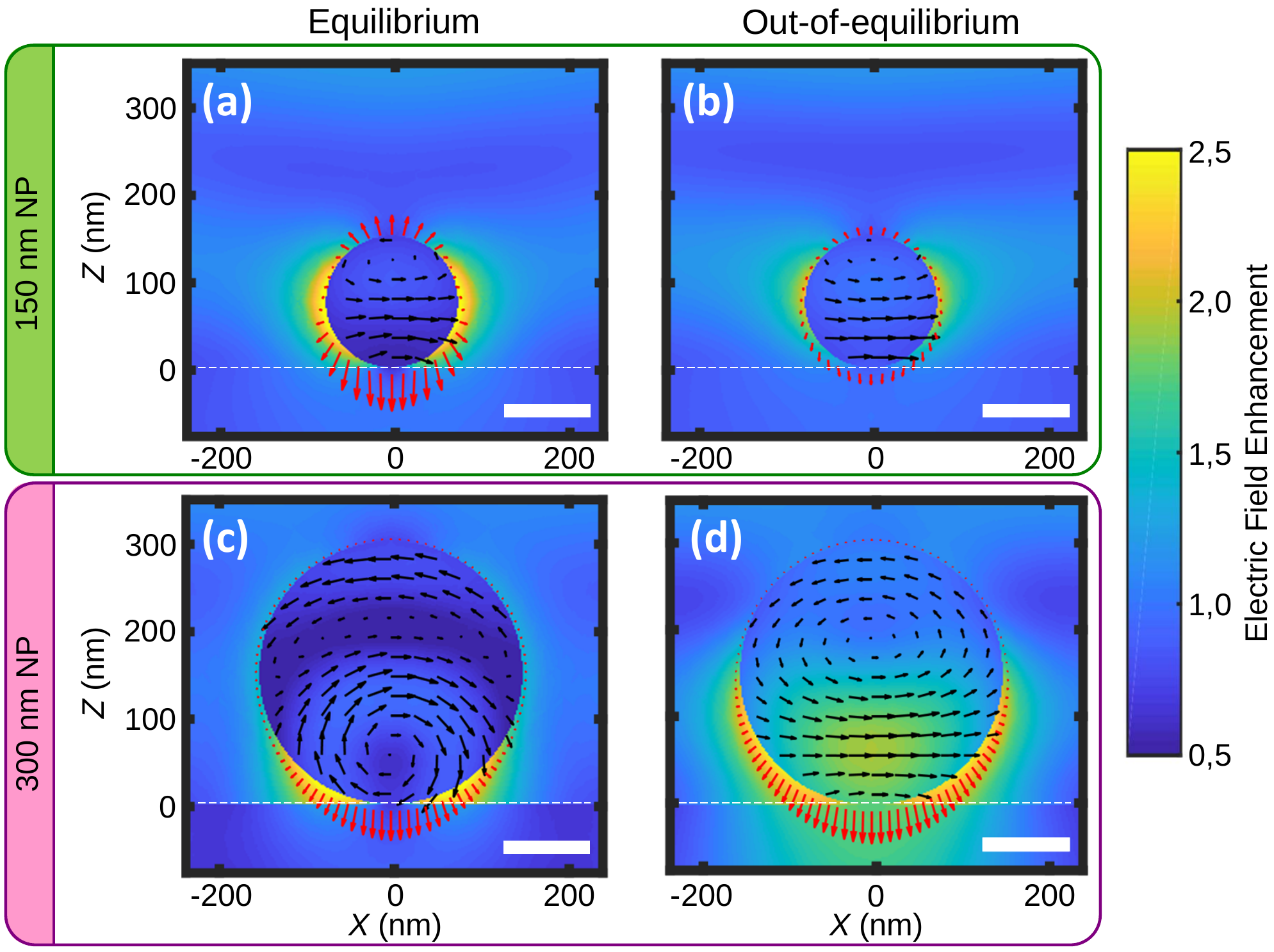}
\caption{\textbf{Electric Field Enhancement.} Electric field enhancement on the plane of the vertical cross-section of the nanoparticle. The panels show the equilibrium (a, c) and out-of-equilibrium (b, d) $F\!E$ for a single CsPbBr$_3$ NP with diameter of $\SI{150}{}$ (a, b) and $\SI{300}{nm}$ (c, d). The incident monochromatic wave centered at $\SI{2.38}{eV}$ propagates from the top to the bottom along $Z$ axis. The black arrows represent the electric field vector $\vec{E}$ in the NP volume. The red arrows represent the Poynting vector $\vec{S}$ outside the NP surface. The dashed white line represents the substrate surface. The scale bar is $\SI{100}{nm}$.
}
\label{fig:ElectrFieldEnh}
\end{figure*} 
The numerical simulations also reproduce the local electric field enhancement $F\!E$, defined as the ratio between the modulus of the electric field ($\vec{E} \left( x,y,z \right)$) and the amplitude of the of the incident plane wave ($E_0$): $F\!E=\| \vec{E} \left( x,y,z \right) \| /E_0$. Fig. \ref{fig:ElectrFieldEnh} shows equilibrium and out-of-equilibrium $F\!E$ at $\SI{2.38}{eV}$ for a single nanoparticle with $150$ and $\SI{300}{nm}$ diameter size. The black arrows represent the electric field vector for the same excitation parameters used to obtain the plots reported in Fig. \ref{fig:simulations}. For $\SI{150}{nm}$ NPs, the equilibrium resonance (panel (a)) is dominated by the \emph{electric dipole} (ED) mode. The ED spatial pattern is maintained also in out-of-equilibrium conditions (panel (b)). 
On the other hand, the nature of the resonance dramatically changes for $\SI{300}{nm}$ NPs. The electric field distribution in Fig. \ref{fig:ElectrFieldEnh}c suggests that both ED and MD (magnetic dipole) resonances contribute to the electromagnetic response at equilibrium, the latter 
exhibiting a typical field leakage towards the substrate.
The MD contribution in $\SI{300}{nm}$ NPs, for the probe spectral region considered ($\SI{2.20}{}-\SI{2.45}{eV}$), is consistent with the observation\cite{MakarovLPR2017} that MD mode takes place when the relation $\lambda_0/n\approx \bar{\phi}$ is fulfilled, with $\lambda_0/n$ being the light wavelength inside the particle. Indeed, from Ref. \citenum{Makarov2018}, $n=2.1$ at $\hbar \omega=\SI{2.2}{eV}$, which gives $\lambda_0 / n =\SI{270}{nm} \approx\SI{300}{nm}$ (see multipole modes decomposition for NPs of different sizes in Sec. S6). 
In the Fano low-$q$ resonance condition, the photo-excitation strongly alters the mode spatial distribution (panel (d)) by shifting up the MD center of mass and increasing the ED mode contribution at the NP-substrate interface. 

In order to connect the near field solutions to the total scattering, it is instructive to plot the Poynting vector at the NP interface (red arrows). In the Fano high-$q$ resonance condition ($\SI{150}{nm}$ NPs in panels (a) and (b)), photo-excitation leads to a quench of the total scattered energy, corresponding to the positive transmittivity variation measured in the experiments. In contrast, the out-of-equilibrium Poynting vector in Fano low-$q$ resonance NPs ($\SI{300}{nm}$ NPs in panels (c) and (d)) indicates an overall transient scattering enhancement associated to the ED and MD photo-induced change and corresponding to the negative measured transmittivity variation. 

\section*{Conclusions}
In conclusion, we have studied the photo-induced sub-ps optical modulation of the Fano resonance formed by the coupling of an excitonic state with Mie modes in halide perovskite NPs. In particular, we have demonstrated that the ultrafast modification of the optical properties induced by the \textit{band gap renormalization} and \textit{band filling} mechanisms, dramatically depends on the geometry of the NPs. In the low-$q$ resonance Fano condition, the contribution of the two effects on the relative transmittivity variation is completely reversed with respect to the high-$q$ resonant case and to what was previously observed in thin films and bulk materials \cite{Tamming2019}.
Importantly, this is the demonstration of the ultrafast control of the optical response in nanoparticles where Mie resonances are coupled with excitons. In previous studies, the ultrafast all-optical switching in Mie-resonant nanoparticles was carried out with standard semiconductors like amorphous silicon \cite{Makarov2015,Shcherbakov2015,Baranov2016,Shaltout2019} and gallium arsenide \cite{Shcherbakov2017}, for which excitonic effects are negligible at room temperature. Beside the physical effects arising from the coupling between exciton and Mie modes, there is important advantage of high absorption and sharp band-edge in CsPbBr$_3$ perovskites, allowing us to greatly reduce the fluence required to observe ultrafast tuning. CsPbBr$_3$ nanoparticles exhibit an amplitude of the transmittivity modulation that is two order of magnitude larger as compared to that measured in silicon-based Mie resonators under similar photo-excitation conditions \cite{Baranov2016}. Although in this work we focused on CsPbBr$_3$ nanoparticles, the results and modeling discussed simply rely on the geometry-driven overlap between the exciton line and the continuum of Mie resonances. As a consequence, the present results can be promptly extended to a wider class of perovskites-based nanonstructures. The present results offer insights into the photophysics of halide perovskite nanoparticles and provide an additional parameter to control their optical properties at frequencies as high as few THz, with impact on perovskite-based optoelectronic devices, metamaterials and switchable nano-antennas. 

\clearpage
\section*{Methods}
\subsection*{Samples} \label{SI_equilibrium_characterization}
Perovskite NPs are deposited on the ITO and glass substrates according to the following protocol. A solution \textbf{1} consists of lead(II) bromide (PbBr$_2$, 36.7 mg, 0.1 mmol) and cesium(I) bromide (CsBr, 21.2 mg, 0.1 mmol) dissolved in 1 mL of anhydrous dimethyl sulfoxide (DMSO). A solution \textbf{2} consists of polyethylene oxide (PEO, average Mw 70 000, 10 mg) stirred at 300 rpm in 1 ml of DMSO at 70 $\SI{}{\celsius}$ for 10 h. \textbf{1} and \textbf{2} are mixed in 1:3 mass ratio and stirred at 100 $\SI{}{\celsius}$ for 10 min to give a perovskite-polymer ink. The latter is deposited on precleaned glass and ITO substrates by spin-casting at 1500 rpm for 1 min. The deposited films are annealed at 130 $\SI{}{\celsius}$ for 3 min to produce NPs with an average size of 150 nm and at 80 $\SI{}{\celsius}$ for 10 min to obtain NPs with an average size of 300 nm. All the procedures are conducted inside a N$_2$-filled glovebox with both O$_2$ and H$_2$O level not exceeding 1 ppm. Extinction spectra are measured in a UV-vis-NIR spectrophotometer (Shimadzu UV-2600).

\subsection*{Time-Resolved Pump-Probe Spectroscopy}
The optical pump-probe set-up is based on a Yb-laser system  (Pharos, Light Conversion), which delivers $\SI{37.5}{}$ $\mu$J, $\SI{270}{}$ fs pulses at $\SI{1030}{nm}$ central wavelength and 400 kHz repetition rate. A portion of the laser with $\SI{30}{}\mu$J energy is used to pump an optical parametric amplifier (Orpheus-F, Light Conversion). The output signal at $\SI{850}{nm}$ is frequency doubled to obtain a pump pulse at $\SI{425}{nm}$ ($\SI{2.9}{eV}$), bandwidth\cite{PolliCerullo2010} $\Delta \lambda \sim \SI{7}{nm}$, with $\SI{52}{}$ nJ energy per pulse that is focused within $\SI{350}{}\, \mu$m spot size (fwhm). The $54$ $\mu$J/cm$^2$ incident pump fluence is such that the average absorbed photon density is $n_{ph}\simeq \SI{6e19}{cm^{-3}}$ (see Sec. S3 for more details). The other portion of the laser is focused on a sapphire crystal to generate a white-light continuum probe ranging from $\SI{500}{nm}$ to $\SI{1000}{nm}$, which is then reduced to the region $\SI{500}{}$-$\SI{600}{nm}$ ($\SI{2.1}{}$-$\SI{2.5}{eV}$) by a colored bandpass glass filter ($\sim$ $\SI{1}{mW}$). The pump is modulated by a mechanical chopper working at $\SI{2.5}{kHz}$. The pump pulse is delayed in time with respect to the probe pulse by using a linear motorized stage (Physik Instrumente M403.2DG). The pump and the probe beams are orthogonally polarized: the pump is linearly horizontally polarized, whereas the probe is 	vertically polarized. The two beams are non-collinearly focused onto the sample: the pump impinges on the sample at normal incidence, while the incident angle of the probe is less than $\ang{10}$. After the interaction with the sample, the linear and differential transmission spectra (at fixed pump-probe delay) are detected by means of a common-path birefringent interferometer, GEMINI (from NIREOS), which generates two collinear replicas of the incoming light. When the relative delay between the two probe replicas is varied, the light interferogram is obtained on a single-pixel photodiode. \\
The linear transmission spectrum is obtained by computing the Fourier Transform of the interferogram, while the differential transmission spectrum is retrieved by computing the Fourier Transform of the interferogram demodulated by a lock-in amplifier at the chopping frequency. The pump-induced relative transmittivity variation is defined as $\delta T/T(\hbar \omega, \Delta t)=[T_{out}(\hbar \omega,\Delta t)-T_{eq}(\hbar \omega)]/T_{eq}(\hbar \omega)$, where $T_{out}$ and $T_{eq}$ are the out-of- and equilibrium transmittivities.

\subsection*{Simulations}
We performed full-vectorial numerical simulations implemented with the finite element method in COMSOL. In our simulations, we consider the linear scattering problem of an electric field impinging on isolated CsPbBr$_3$ nanospheres deposited on a quartz substrate. From the simulations, we calculated the absorption and scattering cross-section of the single nanoparticle, $C_{\tiny{\mbox{abs}}}$ and $C_{\tiny{\mbox{sca}}}$ respectively, to obtain the extinction cross-section $C_{\tiny{\mbox{ext}}}$ of the single nanoparticle as $C_{\tiny{\mbox{ext}}}=C_{\tiny{\mbox{abs}}}+C_{\tiny{\mbox{sca}}}$\cite{bohren2008absorption}. The absorption (scattering) cross-section is calculated as $C_{\tiny{\mbox{abs}}}=w_{\tiny{\mbox{abs}}}/I_i$ ($C_{\tiny{\mbox{sca}}}=w_{\tiny{\mbox{sca}}}/I_i$), where $w_{\tiny{\mbox{abs}}}$ ($w_{\tiny{\mbox{sca}}}$) is the rate at which the energy is absorbed (scattered) by the particle and $I_i$ is the incident light intensity\cite{bohren2008absorption}. The incident field is a linearly polarized plane wave since the focal spot size in the experiment is much larger than the sphere diameter. To obtain the transmittivity of the layer from $C_{\tiny{\mbox{ext}}}$, we assumed that the nanoparticles are weakly coupled and that they are uniformly distributed on the substrate surface with a coverage distribution density $\sigma_{cov}$=3 NP/$\mu$m$^2$. The transmittivity is given by  $T=10^{-X}$, where the extinction $X$ is related to $C_{\tiny{\mbox{ext}}}$ through $X=\sigma_{cov} \cdot C_{\tiny{\mbox{ext}}} \cdot \log_{10}e$\cite{bohren2008absorption}. The equilibrium transmittivity of the nanoparticle is obtained by solving the linear scattering problem for a single CsPbBr$_3$ sphere, where the equilibrium complex refractive index equals the experimental refractive index of CsPbBr$_3$ thin films, reported in Ref.\citenum{Makarov2018}. The out-of-equilibrium transmittivity is obtained by solving the linear scattering problem for a single CsPbBr$_3$ sphere, where the equilibrium complex refractive index is modified according to Eqs. (\ref{eqn:BGR_absorption1}) and (\ref{eqn:absorption_variation1}). This procedure is applied to nanoparticles with diameter of 150 nm and 300 nm. 

\section*{Acknowledgments}
P.F. and C.G, acknowledge financial support from MIUR through PRIN 2015 (Prot. 2015C5SEJJ001) and PRIN 2017 (Prot. 20172H2SC4) programmes. G.F., S.P. and C.G. acknowledge support from Universit\`a Cattolica del Sacro Cuore through D.1, D.2.2, and D.3.1 grants. L.C. acknowledges STARS StG project PULSAR.
The samples preparation was supported by the Russian Ministry of Science and Higher Education (Project 14.Y26.31.0010). F.B. acknowledges financial support from the IDEXLYON Project-Programme Investissements d'Avenir (ANR-16 -IDEX-0005), France. NIREOS acknowledges financial support from the European Union's Horizon 2020 research and innovation programme under grant agreement No 814492 (SimDOME).

\newpage
\bibliographystyle{unsrt}
\bibliography{fileBibliography}

\begin{thebibliography}{10}

\bibitem{review10years}
{A Decade of Perovskite Photovoltaics}.
\newblock {\em Nat. Energy}, 4(1):1, 2019.

\bibitem{Fu2019}
Yongping Fu, Haiming Zhu, Jie Chen, Matthew~P Hautzinger, X.-Y. Zhu, and Song
  Jin.
\newblock {Metal Halide Perovskite Nanostructures for Optoelectronic
  Applications and the Study of Physical Properties}.
\newblock {\em Nat. Rev. Mater.}, 4(3):169--188, 2019.

\bibitem{Green2014}
Martin~A. Green and Henry~J. Ho-Baillie, Anita~Snaith.
\newblock The emergence of perovskite solar cells.
\newblock {\em Nat. Photonics}, 8:506, 2014.

\bibitem{Jang2015}
Dong~Myung Jang, Kidong Park, Duk~Hwan Kim, Jeunghee Park, Fazel Shojaei,
  Hong~Seok Kang, Jae-Pyung Ahn, Jong~Woon Lee, and Jae~Kyu Song.
\newblock Reversible halide exchange reaction of organometal trihalide
  perovskite colloidal nanocrystals for full-range band gap tuning.
\newblock {\em Nano Lett.}, 15(8):5191--5199, 2015.

\bibitem{Protescu2015}
Loredana Protesescu, Sergii Yakunin, Maryna~I. Bodnarchuk, Franziska Krieg,
  Riccarda Caputo, Christopher~H. Hendon, Ruo~Xi Yang, Aron Walsh, and
  Maksym~V. Kovalenko.
\newblock Nanocrystals of cesium lead halide perovskites
  ($\mbox{C}$s$\mbox{P}$b$\mbox{X}$$_3$, $\mbox{X}=\mbox{Cl}$, $\mbox{Br}$ and
  $\mbox{I}$): Novel optoelectronic materials showing bright emission with wide
  color gamut.
\newblock {\em Nano Lett.}, 15(6):3692--3696, 2015.

\bibitem{Byun2017}
Hye~Ryung Byun, Dae~Young Park, Hye~Min Oh, Gon Namkoong, and Mun~Seok Jeong.
\newblock Light soaking phenomena in organic - inorganic mixed halide
  perovskite single crystals.
\newblock {\em ACS Photonics}, 4(11):2813--2820, 2017.

\bibitem{Wehrenfennig2014}
Christian Wehrenfennig, Giles~E. Eperon, Michael~B. Johnston, Henry~J. Snaith,
  and Laura~M. Herz.
\newblock High charge carrier mobilities and lifetimes in organolead trihalide
  perovskites.
\newblock {\em Adv. Mater.}, 26(10):1584--1589, 2014.

\bibitem{Dong-science-2015}
Qingfeng Dong, Yanjun Fang, Yuchuan Shao, Padhraic Mulligan, Jie Qiu, Lei Cao,
  and Jinsong Huang.
\newblock {Electron-Hole Diffusion Lengths $>$ 175 $\mu$m in Solution-Grown
  CH$_3$NH$_3$PbI$_3$ Single Crystals}.
\newblock {\em Science}, 347(6225):967--970, 2015.

\bibitem{yang-science-2015}
Woon~Seok Yang, Jun~Hong Noh, Nam~Joong Jeon, Young~Chan Kim, Seungchan Ryu,
  Jangwon Seo, and Sang~Il Seok.
\newblock High-performance photovoltaic perovskite layers fabricated through
  intramolecular exchange.
\newblock {\em Science}, 348(6240):1234--1237, 2015.

\bibitem{Manser2014}
Joseph~S. Manser and Prashant~V. Kamat.
\newblock Band filling with free charge carriers in organometal halide
  perovskites.
\newblock {\em Nat. Photonics}, 8:737, 2014.

\bibitem{Sheng2015}
ChuanXiang Sheng, Chuang Zhang, Yaxin Zhai, Kamil Mielczarek, Weiwei Wang,
  Wanli Ma, Anvar Zakhidov, and Z.~Valy Vardeny.
\newblock Exciton \textit{versus} free carrier photogeneration in organometal
  trihalide perovskites probed by broadband ultrafast polarization memory
  dynamics.
\newblock {\em Phys. Rev. Lett.}, 114:116601, Mar 2015.

\bibitem{Milot2015}
Rebecca~L. Milot, Giles~E. Eperon, Henry~J. Snaith, Michael~B. Johnston, and
  Laura~M. Herz.
\newblock {Temperature-Dependent Charge-Carrier Dynamics in CH$_3$NH$_3$PbI$_3$
  Perovskite Thin Films}.
\newblock {\em Adv. Funct. Mater.}, 25(39):6218--6227, 2015.

\bibitem{Yang2015}
Ye~Yang, David~P. Ostrowski, Ryan~M. France, Kai Zhu, Jao van~de Lagemaat,
  Joseph~M. Luther, and Matthew~C. Beard.
\newblock Observation of a hot-phonon bottleneck in lead-iodide perovskites.
\newblock {\em Nat. Photonics}, 10:53--59, 2015.

\bibitem{Price2015}
Michael~B. Price, Justinas Butkus, Aditya Jellicoe, Tom C.and~Sadhanala, Anouk
  Briane, Jonathan~E. Halpert, Katharina Broch, Justin~M. Hodgkiss, Justin~M.
  Hodgkiss, Richard~H. Friend, and Felix Deschler.
\newblock Hot-carrier cooling and photoinduced refractive index changes in
  organic-inorganic lead halide.
\newblock {\em Nat. Comm.}, 6 I:8420, 2015.

\bibitem{Herz2016}
Laura~M. Herz.
\newblock Charge-carrier dynamics in organic - inorganic metal halide
  perovskites.
\newblock {\em Annu. Rev. of Phys. Chem.}, 67(1):65--89, 2016.

\bibitem{Tamming2019}
Ronnie~R. Tamming, Justinas Butkus, Michael~B. Price, Parth Vashishtha, Shyamal
  K.~K. Prasad, Jonathan~E. Halpert, Kai Chen, and Justin~M. Hodgkiss.
\newblock Ultrafast spectrally resolved photoinduced complex refractive index
  changes in $\mbox{C}$s$\mbox{P}$b$\mbox{X}$$_3$ perovskites.
\newblock {\em ACS Photonics}, 6(2):345--350, 2019.

\bibitem{Palmieri2020}
Tania Palmieri, Edoardo Baldini, Alexander Steinhoff, Ana Akrap, M{\'{a}}rton
  Koll{\'{a}}r, Endre Horv{\'{a}}th, L{\'{a}}szl{\'{o}} Forr{\'{o}}, Frank
  Jahnke, and Majed Chergui.
\newblock {Mahan Excitons in Room-Temperature Methylammonium Lead Bromide
  Perovskites}.
\newblock {\em Nat. Comm.}, 11(1):850, 2020.

\bibitem{Gholipour2017}
Behrad Gholipour, Giorgio Adamo, Daniele Cortecchia, Harish N.~S.
  Krishnamoorthy, Muhammad.~D. Birowosuto, Nikolay~I. Zheludev, and Cesare
  Soci.
\newblock Organometallic perovskite metasurfaces.
\newblock {\em Adv. Mater.}, 29(9):1604268, 2017.

\bibitem{Makarov2017}
Sergey~V. Makarov, Valentin Milichko, Elena~V. Ushakova, Mikhail Omelyanovich,
  Andrea Cerdan~Pasaran, Ross Haroldson, Balasubramaniam Balachandran, Honglei
  Wang, Walter Hu, Yuri~S. Kivshar, and Anvar~A. Zakhidov.
\newblock Multifold emission enhancement in nanoimprinted hybrid perovskite
  metasurfaces.
\newblock {\em ACS Photonics}, 4(4):728--735, 2017.

\bibitem{Gao2018}
Yisheng Gao, Can Huang, Chenglong Hao, Shang Sun, Lei Zhang, Chen Zhang,
  Zonghui Duan, Kaiyang Wang, Zhongwei Jin, Nan Zhang, Alexander~V. Kildishev,
  Cheng-Wei Qiu, Qinghai Song, and Shumin Xiao.
\newblock Lead halide perovskite nanostructures for dynamic color display.
\newblock {\em ACS Nano}, 12(9):8847--8854, 2018.

\bibitem{Berestennikov2019}
Alexander~S. Berestennikov, Pavel~M. Voroshilov, Sergey~V. Makarov, and Yuri~S.
  Kivshar.
\newblock Active meta-optics and nanophotonics with halide perovskites.
\newblock {\em Appl. Phys. Rev.}, 6(3):031307, 2019.

\bibitem{Tiguntseva2018}
E.~Y. Tiguntseva, G.~P. Zograf, F.~E. Komissarenko, D.~A. Zuev, A.~A. Zakhidov,
  S.~V. Makarov, and Yuri.~S. Kivshar.
\newblock Light-emitting halide perovskite nanoantennas.
\newblock {\em Nano Lett.}, 18(2):1185--1190, 2018.

\bibitem{Makarov2019}
Sergey Makarov, Aleksandra Furasova, Ekaterina Tiguntseva, Andreas Hemmetter,
  Alexander Berestennikov, Anatoly Pushkarev, Anvar Zakhidov, and Yuri Kivshar.
\newblock Halide-perovskite resonant nanophotonics.
\newblock {\em Adv. Opt. Mater.}, 7(1):1800784, 2019.

\bibitem{Manjappa2017}
Manukumara Manjappa, Yogesh~Kumar Srivastava, Ankur Solanki, Abhishek Kumar,
  Tze~Chien Sum, and Ranjan Singh.
\newblock Hybrid lead halide perovskites for ultrasensitive photoactive
  switching in terahertz metamaterial devices.
\newblock {\em Adv. Mater.}, 29(32):1605881, 2017.

\bibitem{Chanana2018}
Ashish Chanana, Xiaojie Liu, Chuang Zhang, Zeev~Valy Vardeny, and Ajay Nahata.
\newblock Ultrafast frequency-agile terahertz devices using methylammonium lead
  halide perovskites.
\newblock {\em Sci. Adv.}, 4(5), 2018.

\bibitem{Huang1018}
Can Huang, Chen Zhang, Shumin Xiao, Yuhan Wang, Yubin Fan, Yilin Liu, Nan
  Zhang, Geyang Qu, Hongjun Ji, Jiecai Han, Li~Ge, Yuri Kivshar, and Qinghai
  Song.
\newblock Ultrafast control of vortex microlasers.
\newblock {\em Science}, 367(6481):1018--1021, 2020.

\bibitem{Makarov2018}
Ekaterina~Y. Tiguntseva, Denis~G. Baranov, Anatoly~P. Pushkarev, Battulga
  Munkhbat, Filipp Komissarenko, Marius Franckevicius, Anvar~A. Zakhidov, Timur
  Shegai, Yuri~S. Kivshar, and Sergey~V. Makarov.
\newblock Tunable hybrid fano resonances in halide perovskite nanoparticles.
\newblock {\em Nano Lett.}, 18(9):5522--5529, 2018.

\bibitem{Shcherbakov2015}
Maxim~R Shcherbakov, Polina~P Vabishchevich, Alexander~S Shorokhov, Katie~E
  Chong, Duk-Yong Choi, Isabelle Staude, Andrey~E Miroshnichenko, Dragomir~N
  Neshev, Andrey~A Fedyanin, and Yuri~S Kivshar.
\newblock {Ultrafast All-Optical Switching with Magnetic Resonances in
  Nonlinear Dielectric Nanostructures}.
\newblock {\em Nano Lett.}, 15(10):6985--6990, oct 2015.

\bibitem{Maragkou2015}
Maria Maragkou.
\newblock {Ultrafast Responses}.
\newblock {\em Nat. Mater.}, 14(11):1086, 2015.

\bibitem{Soukoulis2011}
Costas~M Soukoulis and Martin Wegener.
\newblock {Past Achievements and Future Challenges in the Development of
  Three-Dimensional Photonic Metamaterials}.
\newblock {\em Nat. Photonics}, 5(9):523--530, 2011.

\bibitem{Peruch2017}
Silvia Peruch, Andres Neira, Gregory~A. Wurtz, Brian Wells, Viktor~A.
  Podolskiy, and Anatoly~V. Zayats.
\newblock Geometry defines ultrafast hot-carrier dynamics and kerr nonlinearity
  in plasmonic metamaterial waveguides and cavities.
\newblock {\em Adv. Opt. Mater.}, 5(15):1700299, 2017.

\bibitem{Bennett1990}
B.~R. Bennett, R.~A. Soref, and J.~A. Del~Alamo.
\newblock Carrier-induced change in refractive index of $\mbox{I}$n$\mbox{P}$,
  $\mbox{G}$a$\mbox{A}$s and $\mbox{I}$n$\mbox{G}$a$\mbox{A}$s$\mbox{P}$.
\newblock {\em IEEE J. Quantum Electron.}, 26:113--122, 1990.

\bibitem{Burstein1954}
Elias Burstein.
\newblock Anomalous optical absorption limit in $\mbox{I}$n$\mbox{S}$b.
\newblock {\em Phys. Rev.}, 93:632--633, 1954.

\bibitem{Moss1954}
T.S. Moss.
\newblock The interpretation of the properties of indium antimonide.
\newblock {\em Proc. Phys. Soc., B}, 67(306):775--782, 1954.

\bibitem{Preda2016}
Fabrizio Preda, Vikas Kumar, Francesco Crisafi, Diana Gisell~Figueroa del
  Valle, Giulio Cerullo, and Dario Polli.
\newblock Broadband pump-probe spectroscopy at 20-$\mbox{MH}$z modulation
  frequency.
\newblock {\em Opt. Lett.}, 41:2970--2973, 2016.

\bibitem{Preda2017}
F.~{Preda}, A.~{Oriana}, J.~{Rehault}, L.~{Lombardi}, A.~C. {Ferrari},
  G.~{Cerullo}, and D.~{Polli}.
\newblock Linear and nonlinear spectroscopy by a common-path birefringent
  interferometer.
\newblock {\em IEEE J. Sel. Top. Quant.}, 23(3):88--96, 2017.

\bibitem{Battie2014}
Y.~Battie, A.~Resano-Garcia, N.~Chaoui, Y.~Zhang, and A.~En~Naciri.
\newblock Extended maxwell-garnett-mie formulation applied to size dispersion
  of metallic nanoparticles embedded in host liquid matrix.
\newblock {\em J. Chem. Phys.}, 140(4):044705, 2014.

\bibitem{markovich2014magnetic}
Dmitry~L Markovich, Pavel Ginzburg, AK~Samusev, Pavel~A Belov, and Anatoly~V
  Zayats.
\newblock Magnetic dipole radiation tailored by bubstrates: Numerical
  investigation.
\newblock {\em Opt. Express}, 22(9):10693--10702, 2014.

\bibitem{Richter2017}
Johannes~M Richter, Federico Branchi, Franco {Valduga de Almeida Camargo},
  Baodan Zhao, Richard~H Friend, Giulio Cerullo, and Felix Deschler.
\newblock {Ultrafast Carrier Thermalization in Lead Iodide Perovskite Probed
  with Two-Dimensional Electronic Spectroscopy}.
\newblock {\em Nat. Comm.}, 8(1):376, 2017.

\bibitem{MakarovLPR2017}
Sergey~V. Makarov, Anastasia~S. Zalogina, Mohammad Tajik, Dmitry~A. Zuev,
  Mikhail~V. Rybin, Aleksandr~A. Kuchmizhak, Saulius Juodkazis, and Yuri
  Kivshar.
\newblock Light-induced tuning and reconfiguration of nanophotonic structures.
\newblock {\em Laser Photonics Rev.}, 11(5):1700108, 2017.

\bibitem{Makarov2015}
Sergey Makarov, Sergey Kudryashov, Ivan Mukhin, Alexey Mozharov, Valentin
  Milichko, Alexander Krasnok, and Pavel Belov.
\newblock {Tuning of Magnetic Optical Response in a Dielectric Nanoparticle by
  Ultrafast Photoexcitation of Dense Electron-Hole Plasma}.
\newblock {\em Nano Lett.}, 15(9):6187--6192, sep 2015.

\bibitem{Baranov2016}
Denis~G Baranov, Sergey~V Makarov, Valentin~A Milichko, Sergey~I Kudryashov,
  Alexander~E Krasnok, and Pavel~A Belov.
\newblock {Nonlinear Transient Dynamics of Photoexcited Resonant Silicon
  Nanostructures}.
\newblock {\em ACS Photonics}, 3(9):1546--1551, sep 2016.

\bibitem{Shaltout2019}
Amr~M. Shaltout, Konstantinos~G. Lagoudakis, Jorik van~de Groep, Soo~Jin Kim,
  Jelena Vu{\v c}kovi{\'c}, Vladimir~M. Shalaev, and Mark~L. Brongersma.
\newblock Spatiotemporal light control with frequency-gradient metasurfaces.
\newblock {\em Science}, 365(6451):374--377, 2019.

\bibitem{Shcherbakov2017}
Maxim~R. Shcherbakov, Sheng Liu, Varvara~V. Zubyuk, Aleksandr Vaskin, Polina~P.
  Vabishchevich, Gordon Keeler, Thomas Pertsch, Tatyana~V. Dolgova, Isabelle
  Staude, Igal Brener, and Andrey~A. Fedyanin.
\newblock {Ultrafast All-Optical Tuning of Direct-Gap Semiconductor
  Metasurfaces}.
\newblock {\em Nat. Comm.}, 8(1):17, 2017.

\bibitem{PolliCerullo2010}
D.~Polli, D.~Brida, S.~Mukamel, G.~Lanzani, and G.~Cerullo.
\newblock Effective temporal resolution in pump-probe spectroscopy with
  strongly chirped pulses.
\newblock {\em Phys. Rev. A}, 82:053809, Nov 2010.

\bibitem{bohren2008absorption}
Craig~F Bohren and Donald~R Huffman.
\newblock {\em Absorption and scattering of light by small particles}.
\newblock John Wiley \& Sons, New York, 1 edition, 1983.

\bibitem{Note1}
The absorption coefficient $\alpha $ is related to the dielectric function
  $\varepsilon =\varepsilon ^{(1)}+i \varepsilon ^{(2)}$ via $$ \alpha
  =\protect \frac {2\omega }{c} \protect \tmspace +\thinmuskip {.1667em}
  \protect \sqrt {\protect \frac {-{\varepsilon ^{(1)}}+\protect \sqrt {{\left
  ( \varepsilon ^{(1)} \right )}^2+{\left ( \varepsilon ^{(2)} \right
  )}^2}}{2}}, $$ where $\omega $ is the optical frequency and $c$ the velocity
  of light in vacuum.

\bibitem{Stern1964}
Frank Stern.
\newblock Dispersion of the index of refraction near the absorption edge of
  semiconductors.
\newblock {\em Phys. Rev.}, 133:A1653--A1664, 1964.

\bibitem{Ahmad2017}
Murad Ahmad, Gul Rehman, Liaqat Ali, M.~Shafiq, R.~Iqbal, Rashid Ahmad,
  Tahirzeb Khan, S.~Jalali-Asadabadi, Muhammad Maqbool, and Iftikhar Ahmad.
\newblock Structural, electronic and optical properties of
  $\mbox{C}$s$\mbox{P}$b$\mbox{X}$$_3$ ($\mbox{X}=\mbox{Cl}$, $\mbox{Br}$,
  $\mbox{I}$) for energy storage and hybrid solar cell applications.
\newblock {\em J. Alloy Compd.}, 705:828 -- 839, 2017.

\bibitem{Nilsson1978}
N.~G. Nilsson.
\newblock Empirical approximations for the fermi energy in a semiconductor with
  parabolic bands.
\newblock {\em Appl. Phys. Lett.}, 33(7):653--654, 1978.

\end{thebibliography}

\newpage
\section*{\huge{SUPPLEMENTARY INFORMATION}}



\setcounter{equation}{0}
\renewcommand{\theequation}{S\arabic{equation}}

\setcounter{figure}{0}
\setcounter{section}{0}
\renewcommand{\thefigure}{S\arabic{figure}}
\renewcommand{\thetable}{S\arabic{table}}
\renewcommand{\thesection}{S\arabic{section}}

\section{Supplementary Figure}

\begin{figure*}[h]
\includegraphics[keepaspectratio,clip,width=0.9\textwidth]{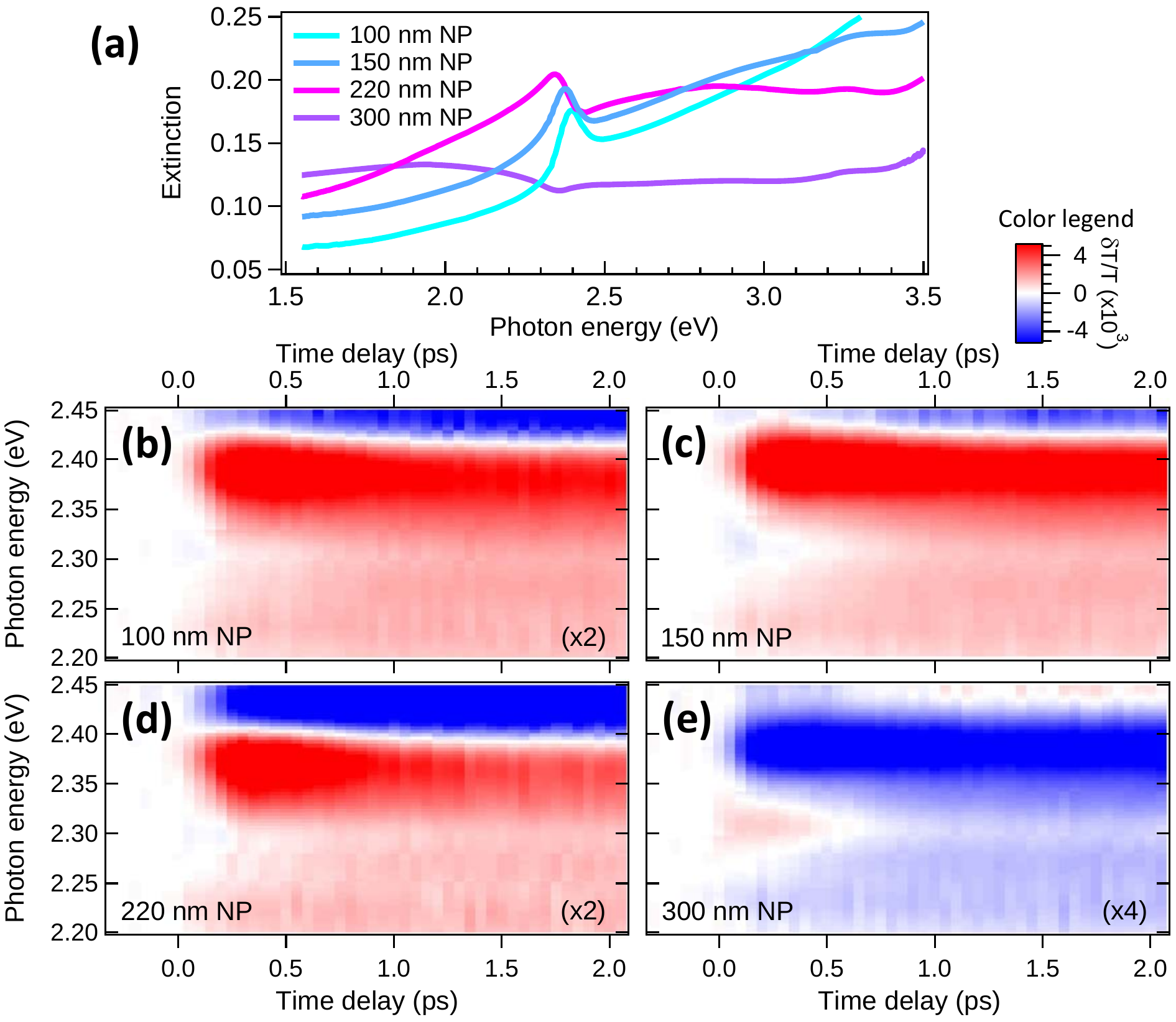}
\caption{ (a) Equilibrium extinction spectra of CsPbBr$_3$ NPs of different size. (b-e) Time- and energy-resolved $\delta T/T$ maps corresponding to the NPs whose extinction is reported in panel (a).
}
\label{fig:mappe2D_suppInfo}
\end{figure*} 

\clearpage
\newpage
\section{Effective medium theory and model for the differential transmission} \label{SI_MG_description}

In Fig. 1 (main text) we show the fit of the Fano profile (Eq. 1 of the main text) to the experimental extinction ($X$) of the $\SI{150}{}$ and $\SI{300}{nm}$ NP samples at equilibrium. {In Eq. 1 of the main text, $X_{\mathrm{bck}}$ accounts for the absorption across the semiconducting edge and its analytical expression is given by a second-order polynomial function
\begin{eqnarray}
X_{\mathrm{bck}}(\hbar\omega)=k_0+k_1 \cdot {\left( {\hbar \omega} \right)}+k_2 \cdot {\left( {\hbar \omega} \right)}^2,
\end{eqnarray}
where $k_0$, $k_1$, and $k_2$ are free parameters.}

Given the relation $X$=$-\log_{10} T$ (where $T$ is the sample transmission), it follows that the differential transmission ($\delta T/T$) is:
\begin{eqnarray} \label{SIeqn:fit_diff_X}
\frac{\delta T}{T}=10^{-\delta X}-1\simeq- \frac{1}{\log_{10} e} \delta X,
\end{eqnarray}
where $\delta X$ is the differential extinction.

\begin{figure*}[t]
\includegraphics[keepaspectratio,clip,width=0.8\textwidth]{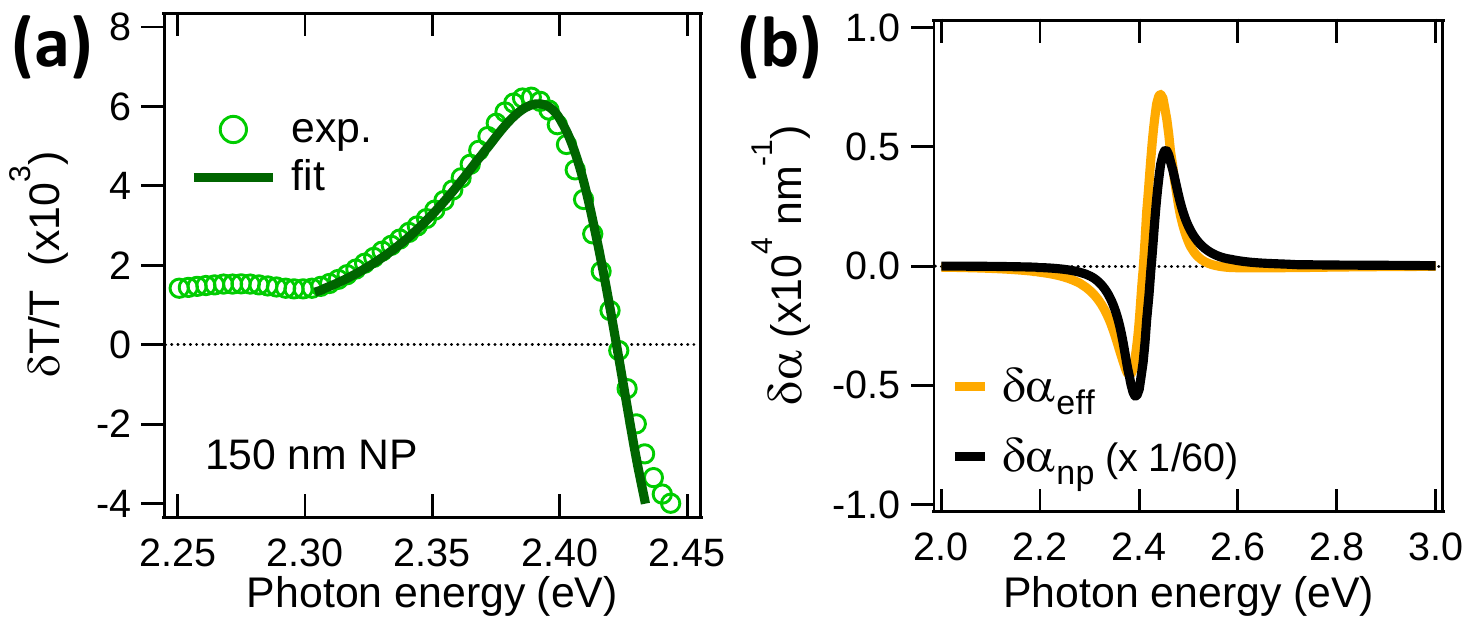}
\caption{\textbf{Out-of-equilibrium properties at long time delays.} (a) Experimental differential transmission spectra (green markers) for 150 nm nanoparticles, taken at $\Delta t=\SI{2}{ps}$ after excitation. The solid line is obtained by fitting Eq. \ref{SIeqn:fit_diff_X} to the experimental data in the energy region near the resonance. (b) Variation of the absorption coefficient of the effective medium $\delta {\alpha_{\tiny{\mbox{eff}}}}$ (yellow solid line) and of the inclusion $\delta {\alpha_{\tiny{\mbox{np}}}}$ (black solid line); the latter is multiplied by a factor $1/60$.
}
\label{fig:tempi_lunghi}
\end{figure*}

Fig. \ref{fig:tempi_lunghi}a displays the experimental $\delta T/T$ spectra of the $\SI{150}{nm}$ NP sample, taken at delay time $\Delta t=\SI{2}{ps}$. The experimental data (circles) can be reproduced (solid line) by assuming a blue-shift $\delta E_g=(5.1\pm0.4)\, \SI{}{meV}$ of the Fano profile (Eq. 1 of the main text).
As discussed in the main text, the observed blue-shift $\delta E_g$ refers to the whole sample, which can be modelled as an effective medium consisting in CsPbBr$_3$ nanoparticles surrounded by air. 
To compare the measured band-gap shift ($\delta E_g$) to the values reported in literature for thin films of similar perovskite compounds, we adopt the Modified Maxwell-Garnett Mie model (MMGM)~\cite{Battie2014}, which accounts for the geometry dispersion of Mie resonators, to extract a scaling factor $\tilde{C}$. This scaling factor allows us to estimate the intrinsic bandgap shift of individual nanoparticles ($\delta E^{NP}_g$) and compare it to results obtained on thin films.

According to the MMGM model, the dielectric function of the effective medium (${\varepsilon_{\tiny{\mbox{eff}}}}$) is related to that of the nanoparticles (${\varepsilon_{\tiny{\mbox{np}}}}$) and of the surrounding medium (${\varepsilon_{\tiny{\mbox{m}}}}$) through the relation:
\begin{eqnarray} \label{eqn:MGT_formula}
\frac{{\varepsilon_{\tiny{\mbox{eff}}}}-{ \varepsilon_{\tiny{\mbox{m}}} }}{{ \varepsilon_{\tiny{\mbox{eff}}} }+2{\varepsilon_{\tiny{\mbox{m}}}}}=\frac{3i{f_{\tiny{\mbox{vol}}}}}{2} \, {\left( \frac{\lambda_0}{\pi \, \sqrt{\varepsilon_{\tiny{\mbox{m}}}} \, \bar{\phi}} \right)}^3 \, \int_{\phi_{1}}^{\phi_{2}} \mathcal{F}\!\left( \phi \right) \, a_1\!\left( \phi;\lambda_0 \right)\, \mbox{d}\phi
\end{eqnarray}
where ${f_{\tiny{\mbox{vol}}}}$ is the volume fraction of the inclusion (see Sec. \ref{SISubSec:voulme_fraction} for the details of the calculation), $\lambda_0$ is the wavelength in vacuum, $\bar{\phi}$ is the average diameter of the NPs, $\mathcal{F}\!\left( \phi \right)$ is the NPs diameter distribution function, $\phi_{1}$ and $\phi_{2}$ are the lower and upper limits of the size distribution. The term $a_1\!\left( \phi;\lambda \right)$ is the \emph{first electric} (see Sec. \ref{SI_Sec:modes_decomposition} for more details) Mie coefficient given by:
\begin{eqnarray}
a_1\!\left( \phi;\lambda_0 \right)\!=\!\frac{{\sqrt{\varepsilon_{\tiny{\mbox{np}}}}} \, {\psi_{1}\!\left( \pi  \phi  \sqrt{\varepsilon_{\tiny{\mbox{np}}}} / \lambda_0 \right)}\, {\psi_1'\!\left( \pi  \phi  \sqrt{\varepsilon_{\tiny{\mbox{m}}}} / \lambda_0 \right)}-{\sqrt{\varepsilon_{\tiny{\mbox{m}}}}} \, {\psi_1\!\left(  \pi  \phi  \sqrt{\varepsilon_{\tiny{\mbox{m}}}} / \lambda_0 \right)} \, {\psi_{1}'\!\left( \pi  \phi  \sqrt{\varepsilon_{\tiny{\mbox{np}}}} / \lambda_0 \right)}}{{\sqrt{\varepsilon_{\tiny{\mbox{np}}}}} \; {\psi_1\!\left( \pi  \phi  \sqrt{\varepsilon_{\tiny{\mbox{np}}}} / \lambda_0 \right)}\,{\xi_{1}' \! \left( \pi  \phi  \sqrt{\varepsilon_{\tiny{\mbox{m}}}} / \lambda_0 \right)}-{\sqrt{\varepsilon_{\tiny{\mbox{m}}}}} \; {\xi_{1}\!\left( \pi  \phi  \sqrt{\varepsilon_{\tiny{\mbox{m}}}} / \lambda_0 \right)} \, {\psi_{1}'\!\left( \pi  \phi  \sqrt{\varepsilon_{\tiny{\mbox{np}}}} / \lambda_0 \right)}},
\end{eqnarray}
where $\psi$ and $\xi$ are the Riccati - Bessel functions. 
We assume that the equilibrium dielectric function of the nanoparticle ${\varepsilon_{\tiny{\mbox{np}}}}$ equals that of the CsPbBr$_3$ {thin film} (which is extracted from Ref. \citenum{Makarov2018}).
According to the experimental distribution of the NPs diameters reported in Fig. 1 of the main text, we assume a normal distribution of diameters given by: \\
$\mathcal{F}\!\left( \phi \right)={1/\sqrt{2 \pi \sigma_{\phi}^{2}}} \; e^{-{\left( \frac{\phi-\bar{\phi}}{\sqrt{2} \, \sigma_{\phi}} \right)}^2 }$, where $\sigma_{\phi}$ is the standard deviation of the distribution.
In the case of the $\SI{150}{nm}$ NPs sample, $\bar{\phi}=\SI{150}{nm}$ and $\sigma_{\phi}=\SI{17}{nm}$.

Within the framework of MMGM theory, we calculated the scaling factor $\tilde{C}$ as the ratio between the absorption\footnote{The absorption coefficient $\alpha$ is related to the dielectric function $\varepsilon=\varepsilon^{(1)}+i \varepsilon^{(2)}$ via
$$
\alpha=\frac{2\omega}{c} \, \sqrt{\frac{-{\varepsilon^{(1)}}+\sqrt{{\left( \varepsilon^{(1)} \right)}^2+{\left( \varepsilon^{(2)} \right)}^2}}{2}},
$$
where $\omega$ is the optical frequency and $c$ the velocity of light in vacuum.} variation of the nanoparticles ($\delta {\alpha_{\tiny{\mbox{np}}}}$) and the absorption variation of the sample ($\delta {\alpha_{\tiny{\mbox{eff}}}}$), \textit{i.e.} $\tilde{C}$=$\delta {\alpha_{\tiny{\mbox{np}}}}$/$\delta {\alpha_{\tiny{\mbox{eff}}}}$. Fig. \ref{fig:tempi_lunghi}b reports the comparison between the variation of the absorption coefficient of the sample $\delta {\alpha_{\tiny{\mbox{eff}}}}$ (yellow solid line) and of the nanoparticles $\delta {\alpha_{\tiny{\mbox{np}}}}$ (black solid line). To calculate $\tilde{C}$ we optimize the overlap between the absorption variations at energies smaller than the exciton peak, which is the region investigated by our pump-probe experiment. Following this procedure we obtain $\tilde{C}$= $\left(  {60 \pm 40}{} \right)$. The error is calculated taking into account the fact that the samples consist in lattices of randomly distributed nanoparticles. The average coverage distribution density is $\sigma_{cov}=\SI{3}{} \; \mbox{NP}/{\mu\mbox{m}}^2$, but it oscillates in the range 1$\div$5 $\mbox{NP}/{\mu\mbox{m}}^2$. 

\newpage
\subsection{Calculation of the volume fraction ${f_{\tiny{\mbox{vol}}}}$}
\label{SISubSec:voulme_fraction}
\begin{figure*}[h]
\includegraphics[keepaspectratio,clip,width=0.5\textwidth]{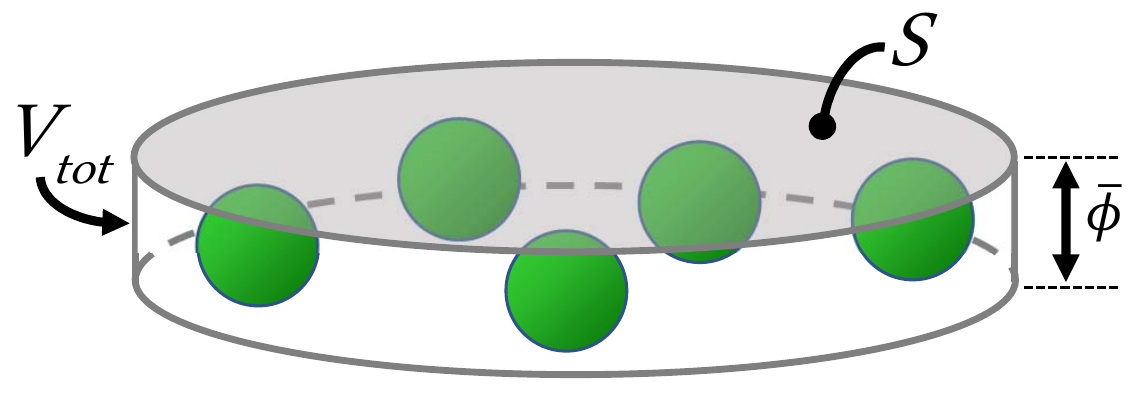}
\caption{\textbf{Calculation of the volume fraction ${f_{\tiny{\mbox{vol}}}}$.} Sketch of the region adopted to estimate the volume fraction. $V_{tot}$ is the volume of the selected region (gray solid line), $S$ is area of the base of the selected region (shadowed gray surface), and $\bar{\phi}$ is mean nanoparticle diameter. The height of the cylinder equals the mean nanoparticle diameter.
}
\label{fig:sketch_f_vol_evaluation}
\end{figure*}

The volume fraction $f_{vol}$ can is estimated as (see Fig. \ref{fig:sketch_f_vol_evaluation}):
$$
f_{vol}=\frac{V \cdot N}{V_{tot}}=\frac{\frac{4}{3} \, \pi \, {\left( \frac{\bar{\phi}}{2} \right)}^3 \cdot \sigma_{cov} \cdot S}{\bar{\phi} \cdot S}=\frac{\pi}{6} \cdot \sigma_{cov} \cdot { {\bar{\phi}} }^2,
$$
where $V$ is the volume of the single nanoparticle, $V_{tot}$ is the volume of the selected region (the height of the cylinder equals the mean nanoparticle diameter), $N$ number of particles within the selected region, $\bar{\phi}$ is the mean nanoparticle diameter, $\sigma_{cov}$ is the average coverage distribution density, and $S$ is area of the base of the selected region. Considering the experimental parameters ${\bar{\phi}}=\SI{150}{nm}$ and $\sigma_{cov}=\SI{3}{} \; \mbox{NP}/{\mu\mbox{m}}^2$, it follows that
$f_{vol}=\SI{0.035}{}$.

\newpage
\section{Calculation of the pump photon density} \label{SI_calculationFreeCarrierDensity}
The initial free-carrier density injected by each pump pulse has been evaluated by taking into account the material extinction, the fluence employed in the pump-probe experiment, and the structural properties of the nanoparticles. The incident fluence is $F=\SI{54}{}\,\mu\mbox{J}$/cm$^2$ at $\lambda=\SI{425}{nm}$. For $\SI{150}{nm}$ NPs, the extinction at the pump wavelength is $X=\SI{0.207}{}$ and the equilibrium transmission results equal to $T=10^{-X}=\SI{0.63}{}$\color{black}. Taking into account the reflection from the substrate ($R$=$\SI{0.08}{}$) and assuming that the energy is absorbed by the nanoparticles, the absorption due to the nanoparticles is $A=1-(R+T)=\SI{0.29}{}$. It follows that the absorbed fluence is $E_{\tiny{\mbox{A}}}=F \cdot A=\SI{15.5}{}\,\mu\mbox{J}$/cm$^2$ and, therefore, the energy absorbed over an area $\mathcal{A}=\SI{1}{} \, \mu {\mbox{m}}^2$ is $E_{\tiny{\mbox{a}}}=E_{\tiny{\mbox{A}}} \cdot \mathcal{A}=\SI{1.55e-13}{J}$. Given an average coverage distribution density of $\sigma_{cov}=\SI{3}{NP}$/$\mu$m$^2$ and assuming a spherical shape for the nanoparticles, which gives a volume $V=\SI{1.767e-15}{cm^{3}}$, the energy absorbed per unit volume is $w_{\tiny{\mbox{a}}}=E_{\tiny{\mbox{a}}}/(3 \cdot V)=\SI{29.2}{J cm^{-3}}=\SI{1.82e20}{eV cm^{-3}}$. Finally, given that the pump is centered at $\lambda=\SI{425}{nm}$, the photon density per unit volume at the pump wavelength is $n_{ph}=\SI{62.6 e18}{cm^{-3}}$.

\newpage

\section{Analysis of the time-resolved traces} \label{analysis_tr}
The time-resolved $\delta  T/T \! \left( \Delta t \right)$ traces, at fixed probe energy, are analysed according to the following double-exponential model. This choice is due to the fact that we want to describe the dynamics of the differential signal as the sum of the contributions due to \emph{bandgap renormalization} (BGR) and \emph{band filling} (BF). The $\delta T/T \! \left( \Delta t \right)$ model is given by the convolution between a gaussian function $\mathcal{G}$, describing the experimental time resolution (given by the pump temporal width), and the material response function $\mathcal{F}$ :
\begin{eqnarray}
\delta  T/T \! \left( \Delta t \right)= \mathcal{G}\!\left( \Delta t \right) * \mathcal{F} \!\left( \Delta t \right),
\qquad \mbox{where} \qquad
\mathcal{G}\!\left( t \right) =\sqrt{\frac{4 \, \ln 2}{\pi \, \tau_p^2}} \, e^{-{\frac{4 \, \ln 2}{\tau_p^2}}\; t^2}
\qquad \mbox{and} \qquad
\end{eqnarray}
\begin{multline} \label{eqn:model_dynamics}
\mathcal{F}\!\left( \Delta t \right) =\theta \! \left( \Delta t-t_0 \right) \cdot  \left[ A_1 \cdot \left( 1-e^{-\frac{\Delta t-t_0}{\tau_{R1}}} \right) \cdot  e^{-\frac{\Delta t-t_0}{\tau_{D1}}}+ \right.\\
\left. + A_2 \cdot  \left( 1-e^{-\frac{\Delta t-t_0}{\tau_{R2}}} \right) \cdot  e^{-\frac{\Delta t-t_0}{\tau_{D2}}}\right].
\end{multline}
In the previous expressions, $\tau_p$=40 fs is the FWHM of the pump-laser pulse, $t_0$ is the zero-time offset, $\tau_{R1}$ is the BGR rise time, $\tau_{D1}$ is the BGR decay time, $\tau_{R2}$ is the BF rise time, $\tau_{D2}$ is the BF decay time, $A_1$  and $A_2$ are the amplitude factors of the two mechanisms.

\begin{table}[h]
\centering

\caption{Parameters extracted from the fit procedure applied to the $\delta T/T$ dynamics at different probe energies}
\begin{tabular}{lcc}
\toprule
{Parameter} & 150-nm-NP & 300-nm-NP \\ 
\midrule
 $\tau_{R1} \;  (\SI{}{fs})$   &  200 $\pm$ 10	 & 	190 $\pm$ 10	\\
$\tau_{D1} \;  (\SI{}{fs}) $   &	410 $\pm$ 10	 &	390 $\pm$ 10	\\
$\tau_{R2} \;  (\SI{}{fs}) $   &	500 $\pm$ 20	 &	460 $\pm$ 20	\\
\bottomrule
\end{tabular}
\label{tab:fit-dynamics-tr-output}
\end{table}

\clearpage
\newpage
\section{Models of the dynamics of the single nanoparticle optical properties} \label{SI_math_mod_BGR_and_BF}

Here, we report the description of the models adopted to establish the role of the different physiscal phenomena in determining the out-of-equilibrium optical properties of the single nanoparticle. Upon free-carriers injection, the transmission variation is controlled by three main components \cite{Bennett1990}: the Drude term (D), the band filling (BF) and the bandgap renormalization (BGR) effects. These processes give rise to a modulation of both the refractive index and absorption, as given by:
\begin{subequations}
\begin{align}
\delta n={\delta n}_{D}+{\delta n}_{BGR}+{\delta n}_{BF}\\
\delta \alpha={\delta \alpha}_{D}+{\delta \alpha}_{BGR}+{\delta \alpha}_{BF}
\end{align}
\end{subequations}
For each process ($i$=D, BGR, BF), the refractive index and absorption variations are constrained by the following Kramers-Kr\"onig relations:
\begin{eqnarray} \label{eqn:KK}
{\delta n}_i\! \left( \hbar \omega; n_{fc} \right)=\frac{2 c \hbar}{e^2} \; \mbox{P.V.}  \int_{0}^{+\infty} \frac{{\delta \alpha}_i\! \left( \xi; n_{fc} \right)}{\xi^2-{\left( \hbar \omega \right)}^2} \, \mbox{d}\xi,
\end{eqnarray}
where $c$ is the speed of light in vacuum, $e$ is the electron charge, $\hbar$ is the Planck's constant, and \emph{P.V.} is the Cauchy principal value \cite{Bennett1990, Stern1964}. 
As it will be described in the following, these three components depend on the pump-injected free-carriers density $n_{fc}$. In our analysis, we assume an injected free-carrier density $n_{fc}=\SI{1.2e20}{cm^{-3}}$ for 150 nm NPs; moreover, the pump-injected free-electron density in the conduction band ($n_{e}$) is assumed equal to the pump-injected free-holes density in the valence band ($n_{h}$), \textit{i.e.} $n_{e}\simeq n_{h}\simeq n_{fc}/2$.

\subsection{Drude}
The Drude term represents the physical mechanism for which the photon-absorption promotes a free-carrier to a higher energy state within the same band. The corresponding change in the refractive index is given by:
\begin{eqnarray}
{\delta n}_{D} \left( E; n_{fc} \right)=-   \frac{n_{fc} \, \hbar^2 \, e^2}{4\, m^{*} \, n_{0} \, \varepsilon_{0} \, \left( E^2+\hbar^2 \, \gamma^2 \right)},
\end{eqnarray}
where $m^{*}={\left( m_{e}^{-1}+m_{h}^{-1} \right)}^{-1}$ is the reduced-effective mass ($m^{*}=0.072 \, m_0$, \cite{Protescu2015}), $n_0$ is the refractive index at equilibrium and $\gamma$ is the inverse collision time of the carriers.

\subsection{Bandgap Renormalization}
In the case of parabolic bands, the optical absorption $\left( \alpha_0 \right)$ of electrons (holes) from the valence (conduction) band to the conduction (valence) band is given by the square-root law:
\begin{eqnarray}\label{eqn:absorption_equilibrium}
\alpha_{eq}\! \left( \hbar \omega; E_g^{0} \right)=
\begin{cases}
0 & \mbox{for} \; \hbar \omega < E_g^{0},\\
\frac{{C_1}}{\hbar \omega} \, \sqrt{\hbar \omega-E_g}+C_2 \, \frac{\Gamma/2}{{\left[ \hbar \omega-\left( E_g^{0}-E_x \right) \right]}^2+{\left( \Gamma/2 \right)}^2} & \mbox{for} \;  \hbar \omega \geq E_g^{0}
\end{cases},
\end{eqnarray}
where $E_g^{0}$, $E_x$, and $C$ are respectively the band-gap energy, the exciton binding energy and a constant. The bandgap renormalization term is modeled as a rigid translation (red shift) of the absorption curve:
\begin{eqnarray} \label{eqn:BGR_absorption}
\delta  \alpha_{BGR} \! \left(\hbar\omega; n_{fc}/n_{cr} \right)=\alpha_{eq}\! \left(\hbar\omega; E^0_g-\delta E_{{BGR}}\! \left(n_{fc}/n_{cr} \right) \right)-\alpha_{eq}\! \left(\hbar\omega; E^0_g \right),
\end{eqnarray} 

where $\delta E_{{BGR}}$ is the free-carrier dependent bandgap shift, whose expression is
\begin{eqnarray}
\delta E_{{BGR}}\! \left(n_{fc}/n_{cr} \right)=
\begin{cases}
 \frac{C_3}{ \varepsilon_{s}} \, {\left( 1 - \frac{n_{fc}}{2 \, n_{cr}}\right)}^{1/3}, & n_{fc}/2 \geq n_{cr} \\
0, & n_{fc}/2 < n_{cr}
\end{cases},
\end{eqnarray}
where $C_3$=0.05 is a fitting parameter,  $\varepsilon_{s}$= 4 (Ref. \citenum{Ahmad2017}) is the relative static dielectric constant, and  $n_{cr}$ is the critical concentration of free carriers \cite{Bennett1990}, which is calculated as
\begin{eqnarray}
n_{cr} \left[ \SI{}{cm^{-3}} \right]= {\left( \frac{m^{*}/m_0}{1.4 \, \varepsilon_{s} } \right)}^{3} \cdot \SI{1.6e24}{}=\SI{3.4e18}{}.
\end{eqnarray}

\subsection{Band Filling}
The band filling term implies a free-carriers induced modulation of the interband absorption for photon energies slightly above the nominal bandgap.  In presence of free-carriers injection, the variation of the intraband absorption is described by the following expression:
\begin{eqnarray}\label{eqn:absorption_variation}
\delta  \alpha_{BF} \! \left( \hbar\omega; n_{fc},T^{*} \right)=
\alpha_{eq}\! \left( \hbar\omega; E^0_g \right) \, \left[ f_{v} \! \left(\hbar\omega;  E^*_{F_{v}}, T^{*} \right)-f_{c} \! \left( \hbar\omega; E^*_{F_{c}}, T^{*} \right) -1 \right],
\end{eqnarray}
where
\begin{eqnarray} \label{eqn:Fermi-Dirac_distributions}
f_{v} \! \left( \hbar\omega; E^*_{F_{v}},T^{*} \right)={\left[1+\exp\left( \frac{E_{a}-E^*_{F_{v}}}{k_{B} T^{*}} \right) \right]}^{-1}
\; \mbox{and} \;
f_{c} \! \left( \hbar\omega; E^*_{F_{c}},T^{*} \right)={\left[1+\exp\left( \frac{E_{b}-E^*_{F_{c}}}{k_{B} T^{*}} \right) \right]}^{-1}
\end{eqnarray}
are the Fermi-Dirac distributions for the electrons in the conduction band and holes in the valence band, respectively. In Eq. \ref{eqn:Fermi-Dirac_distributions}, $T^{*}$ is the effective temperature, $E_{a}$ and $E_{b}$ denote an energy level in the valence and conduction band, and $E^*_{F_{v}}$ and $E^*_{F_{c}}$ are the carrier-dependent quasi-Fermi levels. The effective temperature $T^{*}$ is computed by considering the excess energy of the pump-excited free-carriers which, in our case, corresponds to $T^{*} \approx \SI{3000}{K}$ (we assume that the energy is equally distributed between electrons and holes, \cite{Yang2015}). For a given photon energy $\hbar\omega$, the values of $E_{a}$ and $E_{b}$ are uniquely defined on the basis of energy and momentum conservation:
\begin{eqnarray}
 E_{a}= -\left(\hbar\omega-E_g  \right) \left(  \frac{m_{e}}{m_{e}+m_{h}} \right)-E_g
\quad \mbox{and} \quad
 E_{b} =\left(\hbar\omega-E_g  \right) \left(  \frac{m_{h}}{m_{e}+m_{h}} \right).
\end{eqnarray}
The value of the carrier-dependent quasi-Fermi levels is computed by the Nilsson approximation \cite{Nilsson1978}:
\begin{eqnarray}
\label{eq:F_eff_v}
E^*_{F_{v}}=-\left[ \ln \! \left( \frac{n_{fc}}{2N_{v}} \right)+\frac{n_{fc}}{2N_{v}} {\left[ 64+0.05524 \cdot \frac{n_{fc}}{2N_{v}} \cdot \left( 64+\sqrt{\frac{n_{fc}}{2N_{v}}} \right)  \right]}^{-1/4} \right] \, k_{B} T^{*}-E_g^{0}
\end{eqnarray}
and
\begin{eqnarray}
\label{eq:F_eff_c}
E^*_{F_{c}}=\left[  \ln \! \left( \frac{n_{fc}}{2N_{c}} \right)+\frac{n_{fc}}{2N_{c}} {\left[ 64+0.05524 \cdot \frac{n_{fc}}{2N_{c}} \cdot \left( 64+\sqrt{\frac{n_{fc}}{2N_{c}}} \right)  \right]}^{-1/4}  \right] \, k_{B} T^{*},
\end{eqnarray}
where the \emph{zero} energy level is set at the bottom of the conduction band. $N_{v}$ and $N_{c}$ are the effective density of states in the valence and conduction band, respectively, given by
\begin{eqnarray}
N_{v}=2 \, {\left( \frac{m_{h} \, k_{B}\, T^{*}}{2 \pi \hbar} \right)}^{3/2}
\quad \mbox{and} \quad
N_{c}=2 \, {\left( \frac{m_{e} \, k_{B}\, T^{*}}{2 \pi \hbar} \right)}^{3/2}.
\end{eqnarray}

The refractive index variation due to bandgap renormalization and band filling is obtained after the application of the Kramers-Kr\"onig relations (see Eq. \ref{eqn:KK}). The possible photoinduced variation of interband transitions at high energies, \textit{i.e.} beyond the experimental accessible energy range, would gives rise to an additional refractive index variation, which is not accounted for by Eq. \ref{eqn:absorption_variation}. This contribution is considered by assuming an additional frequency independent refractive index variation, ${\delta n}_0$, which can be adjusted to finely match the ratio between the amplitudes of the $\delta T/T$ signals at the energy $\hbar\omega\simeq$2.4 eV for the 150 nm and 300 nm NPs. 

\newpage
\section{Mie-theory of single particle}

\subsection{Extinction, Scattering, and Absorption}\label{SI_Sec:Mie_theory_efficiency_factors}
\begin{figure*}[h]
\includegraphics[keepaspectratio,clip,width=0.8\textwidth]{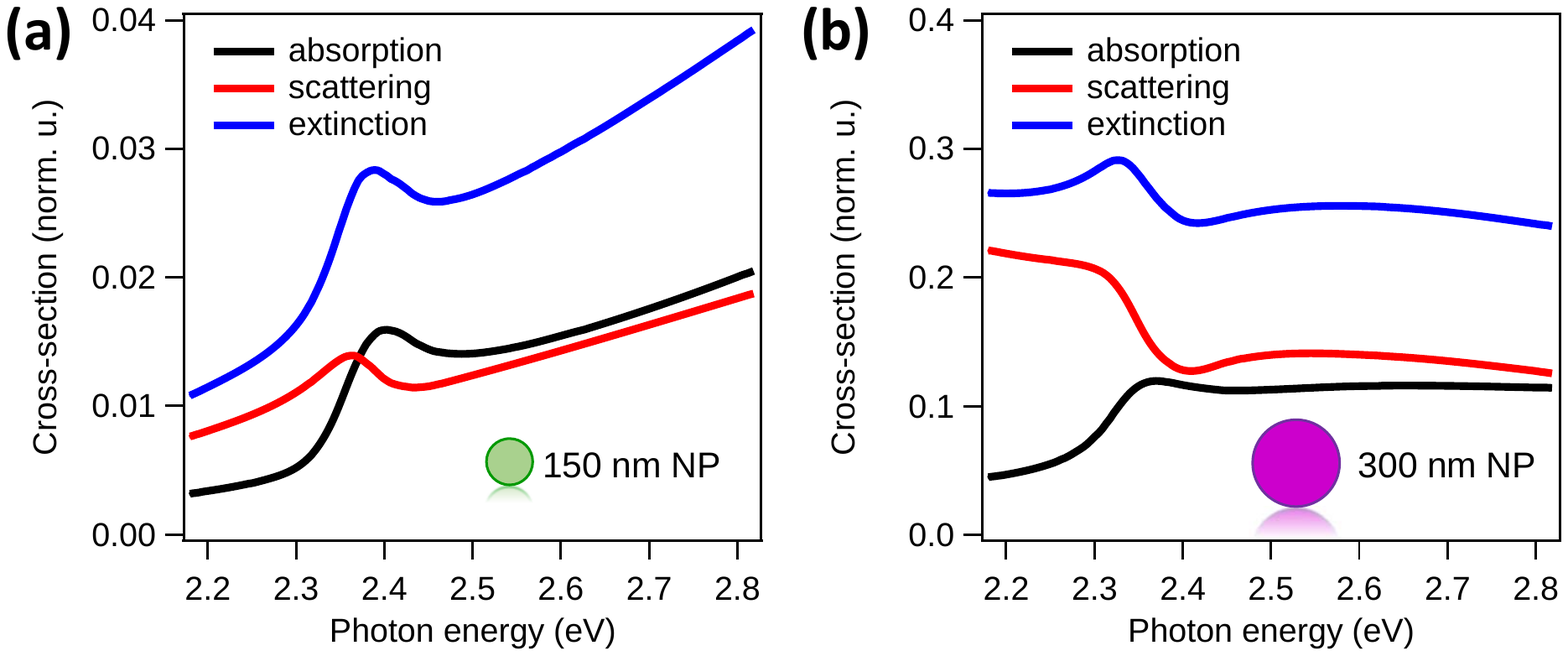}
\caption{\textbf{Cross Sections for spherical CsPbBr$_3$ nanoparticles.}
Extinction, scattering and absorption cross-sections of spherical CsPbBr$_3$ nanoparticles with diameter of 150 (panel a) and 300 nm (panel b) at equilibrium. 
}
\label{SIfig:efficiency_factors}
\end{figure*}

In order to clarify how scattering and absorption mechanisms contribute to total extinction, we analytically calculated the scattering, extinction and absorption cross-sections ($C_{\tiny{\mbox{sca}}}$, $C_{\tiny{\mbox{ext}}}$, and $C_{\tiny{\mbox{abs}}}$, respectively) of a single sperical CsPbBr$_3$ particle, within the framework of Mie theory~\cite{bohren2008absorption}. This model applies to the ideal case of isolated nanoparticles in a surrounding medium, but it is instructive to qualitatively address the role of the scattering and absorption processes. The analytical expressions are given by:
\begin{subequations}\label{SIeqn:crossSections}
\begin{align}
C_{\tiny{\mbox{sca}}} &=  \frac{2\pi}{{\mbox{k}}^2} \sum^{\infty}_{h=1} (2h+1) \; (|a_h|^2+|b_h|^2)\label{eq:46_C_sca}\\
C_{\tiny{\mbox{ext}}} &=  \frac{2\pi}{{\mbox{k}}^2} \sum^{\infty}_{h=1} (2h+1) \; \Re(a_h+b_h)\label{eq:46_C_ext}\\
C_{\tiny{\mbox{abs}}} &=C_{\tiny{\mbox{ext}}}-C_{\tiny{\mbox{sca}}} \label{eq:46_C_abs}
\end{align}
\end{subequations}
where the index $h$ is the multipole order, $a_h$ and $b_h$ are the \emph{scattering Mie-coefficients} of the order $h$, k=$\frac{2\pi \sqrt{\varepsilon_{\tiny{\mbox{m}}}}}{\lambda_0}$ is the wave-number in the surrounding medium, ${\varepsilon_{\tiny{\mbox{m}}}}$ is the dielectric function of the surrounding medium, and $\lambda_0$ is the wavelength in vacuum. Assuming that the permeability of the particle equals to that of the surrounding medium, \emph{Mie coefficients} can be obtained thanks to the following expressions:
\begin{subequations} \label{eq:44}
\begin{align}
a_h &= \frac{m\psi_h(mx)\psi'_h(x) - \psi_h(x)\psi'_h(mx)}{m\psi_h(mx)\xi'_h(x) - \xi_h(x)\psi'_h(mx)}\label{eq:44_ah}\\
b_h &= \frac{\psi_h(mx)\psi'_h(x) - m\psi_h(x)\psi'_h(mx)}{\psi_h(mx)\xi'_h(x) - m\xi_h(x)\psi'_h(mx)},\label{eq:44_bh}
\end{align}
\end{subequations}
where $\psi$ and $\xi$ are the Riccati - Bessel functions, $m=\sqrt{\varepsilon_{\tiny{\mbox{np}}}}/\sqrt{\varepsilon_{\tiny{\mbox{m}}}}$ is the relative refractive index, ${\varepsilon_{\tiny{\mbox{np}}}}$ is the dielectric function of the nanoparticle, $x=\pi \phi \sqrt{\varepsilon_{\tiny{\mbox{m}}}}/\lambda_0$ is the size parameter, and $\phi$ is the sphere diameter.
Basing on these equations and assuming that the dielectric function of a CsPbBr$_3$ nanoparticle equals that of the thin film of the same material (extracted from Ref. \citenum{Makarov2018}), we calculate the cross-sections of the CsPbBr$_3$ spherical NPs with diameter of 150 and 300 nm (see Fig. \ref{SIfig:efficiency_factors}). In both cases, near the exciton region, the scattering, characterized by a Fano asymmetric lineshape, dominates the total extinction.

\subsection{Modes Decomposition (equilibrium and out-of-equilibrium)} \label{SI_Sec:modes_decomposition}
\begin{figure*}[h]
\includegraphics[keepaspectratio,clip,width=0.8\textwidth]{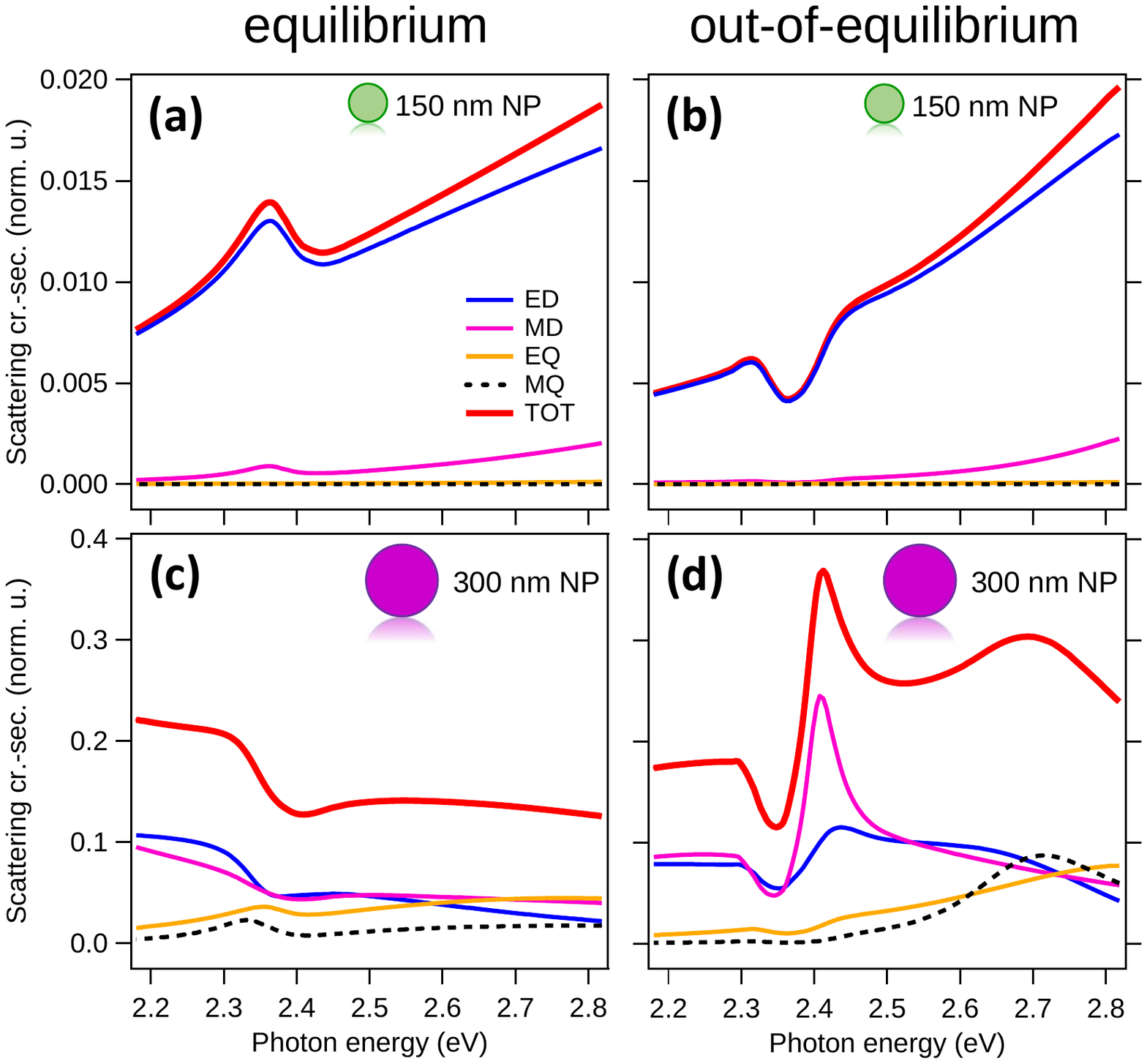}
\caption{\textbf{Modes decomposition for spherical CsPbBr$_3$ nanoparticles.} Calculated scattering cross-section (red solid lines) for a spherical CsPbBr$_3$ nanoparticle of diameters 150~nm (a, b) and 300~nm (c, d), as well as contributions to the four strongest Mie modes in visible spectral range: electric dipole (ED, blue solid lines), magnetic dipole (MD, magenta solid lines), electric quadrupole (EQ, yellow solid lines), and magnetic quadrupole (MQ, black dashed lines). The spectra are calculated for unperturbed material before photo-excitation (a, c), and after photo-excitation assuming  an injected free carrier density $n_{fc} = \SI{1.2e20}{cm^{-3}}$ (b, d). 
}
\label{fig:decomposition}
\end{figure*}

In this section we describe how multipole decomposition of the scattering cross-section is spectrally modified by the photo-excitation process in the simple case of an isolated nanoparticle in a surrounding medium. The outcome of numerical simulations of the full electromagnetic problem in the realistic configuration (nanoparticle+substrate) is discussed in the main text. In expression \ref{SIeqn:crossSections}, $a_h$ (\emph{electric}) and $b_h$ (\emph{magnetic}) are the Mie-coefficients of the order $h$. The multipole modes are named according to the order: $h$=$1$ stands for dipole mode, 2-quadrupole, 3-octupole, \textit{etc.} Fig. \ref{fig:decomposition}a and c exhibit the scattering cross-section  at equilibrium (obtained as described in \ref{SI_Sec:Mie_theory_efficiency_factors}), together with the contribution provided by the four strongest multipole modes (dipolar (D) and quadrupolar (Q) modes of magnetic (M) and electric (E) types), for a CsPbBr$_3$ particle with diameter of 150 and 300 nm (panel (a) and (c), respectively). The out-of-equilibrium scattering cross-section  (\textit{i.e.}, after photo-excitation) is calculated thanks to expressions \ref{eq:46_C_sca}, \ref{eq:44} by using a perturbed perovskite dispersion, $\varepsilon_{\tiny{\mbox{out}}}=\varepsilon_{\tiny{\mbox{np}}}+\delta \varepsilon_{\tiny{\mbox{np}}}$. The photo-induced variation of the nanoparticles dispersion $\delta \varepsilon_{\tiny{\mbox{np}}}$ is obtained as described in Sec. \ref{SI_math_mod_BGR_and_BF}. The out-of-equilibrium cross-section, together with the contribution provided by the four strongest multipole modes, are reported in panels (b) and (d) for a CsPbBr$_3$ particle with diameter of 150 and 300 nm, respectively.

\end{document}